\documentclass[preprint,12pt,authoryear]{elsarticle}
\usepackage[T1]{fontenc}
\usepackage{multirow,rotating}
\usepackage{color}
\usepackage{hyperref}
\usepackage{tikz-cd}
\usepackage{array}
\usepackage{bm}
\usepackage{amsmath,amssymb,amsfonts}
\usepackage{graphicx}
\usepackage{tcolorbox}
\usepackage{mathtools,nccmath}
\usepackage[makeroom]{cancel}

\usepackage{etoolbox, xparse} 
\usepackage{ragged2e}
\usepackage{adjustbox}
\usepackage{rotating}
\usepackage{tikz}
\usepackage{natbib}
\usetikzlibrary{shapes.geometric, arrows}
\usepackage{algorithm}%
\usepackage{algpseudocode}%
\usepackage{algorithmicx}%
\usepackage[algo2e]{algorithm2e}%
\usepackage{soul}
\usepackage{booktabs}
\usepackage{caption}
\usepackage{xr}
\usepackage[normalem]{ulem}
\usepackage{subcaption}
\usepackage{bbm}

\hypersetup{
    pdftitle={INLA-RF: A Hybrid Modeling Strategy for Spatio-Temporal Environmental Data},
    pdfauthor={Mario Figueira, Michela Cameletti, Luca Patelli},
    pdfsubject={Spatio-temporal environmental modeling},
    pdfkeywords={Hybrid models, INLA, Random Forest, SPDE, Bayesian inference},
    colorlinks=true,
    linkcolor=blue,
    citecolor=blue,
    urlcolor=blue
}

\journal{Computational Statistics and Data Analysis}

\begin{document}

\begin{frontmatter}
\title{INLA-RF: A Hybrid Modeling Strategy for Spatio-Temporal Environmental Data}

\author[label1]{Mario Figueira}
\affiliation[label1]{organization={Department of Statistics and Operations Research, University of Valencia},
             city={Valencia},
            country={Spain}}

\author[label2]{Michela Cameletti\corref{cor1}}
\ead{michela.cameletti@unibg.it}
\cortext[cor1]{Corresponding author}
\affiliation[label2]{organization={Department of Economics, University of Bergamo},
  city={Bergamo},
  country={Italy}}
            
\author[label2]{Luca Patelli}


\begin{abstract}
Environmental processes often exhibit complex, non-linear patterns and discontinuities across space and time, posing significant challenges for traditional geostatistical modeling approaches. In this paper, we propose a hybrid modeling strategy that combines the interpretability and principled uncertainty quantification of statistical models with the predictive flexibility of machine learning methods. Specifically, we introduce two novel algorithms, collectively named INLA-RF, which iteratively couple a Bayesian spatio-temporal model - fitted via the INLA-SPDE approach - with Random Forest. The proposed framework also allows uncertainty propagation across modeling stages, an aspect often overlooked in existing hybrid approaches. In addition, we propose an effective stopping criterion based on the Kullback-Leibler divergence.

We assess the predictive performance and uncertainty quantification of the INLA-RF algorithms using two simulation studies and a real air pollution application. Overall, the results indicate that INLA-RF can improve spatio-temporal predictive accuracy while retaining interpretability and providing coherent uncertainty estimates.

\end{abstract}

\begin{keyword}
    Hybrid models \sep INLA \sep  Random Forest \sep SPDE \sep Latent Gaussian Models \sep Uncertainty propagation 
\end{keyword}

\end{frontmatter}


\section{Introduction}

Environmental phenomena are complex. They are usually observed at multiple time points and across various spatial locations, interacting with natural, physical and anthropogenic factors. When analyzing environmental data, the main goal is to understand hidden spatio-temporal patterns and to assess the relationship between the predictors and the response variable. Additionally, predictive models enable prediction at new spatial locations or future time points.

Statistical models, such as Gaussian process regression \citep{StatsforSpatioTemporal2011, SPDEbook, STwithR2019}, are well-established tools widely applied in several environmental contexts, including air pollution prediction, ecological modeling and climate forecasting. These models rely on the likelihood function for inference, ensuring a proper quantification of uncertainty. In particular, within a Bayesian framework, the well-established INLA-SPDE approach\footnote{The acronym INLA-SPDE refers to the Integrated Nested Laplace Approximation approach combined with the Stochastic Partial Differential Equations approach.} \citep{Rue_2009_INLA, lindgren_spde} provides a powerful framework based on parametric Latent Gaussian Models (LGM), enabling efficient and scalable inference while capturing spatio-temporal dependencies.

While statistical models are inherently interpretable, they may lack the flexibility needed to capture the complex, non-linear patterns and discontinuities that usually characterize environmental phenomena \citep{2023MLDLecology, statiticalDL2023, 2024REVIEWatmos}. In contrast, Machine Learning (ML) and Deep Learning (DL) models are particularly well-suited for modeling non-linear patterns and managing high-dimensional data. A key advantage of these methods is that they do not require to specify a functional form for the features-output relationship, as they learn directly from the data. However, they typically offer limited interpretability and lack native mechanisms for uncertainty quantification. Moreover, they typically capture spatio-temporal dependence only implicitly through covariates, rather than through an explicit spatio-temporal covariance function.

Hybrid models, which integrate statistical models with ML/DL methods, have recently emerged as a powerful approach \citep{statiticalDL2023}. These models combine the strengths of both paradigms: statistical methods offer interpretability and uncertainty quantification, while the flexibility of ML/DL methods can help in capturing complex non-linear relationships and high-dimensional dependencies. This synergy makes hybrid models particularly well-suited for modeling complex environmental data, enhancing both the predictive capacity and the ability to uncover complex spatio-temporal patterns.
%
%
Existing hybrid approaches combining statistical and ML/DL models can be broadly grouped into three families. 
First, \emph{residual learning} (or two-stage modeling) fits two models sequentially, with the second model trained on the residuals from the first to capture remaining structure; depending on the modeling objective, either the statistical model or the ML/DL model may be used in the first stage \citep[e.g.,][]{comparinglargedataset2020, deMattosNeto2022, Niraula2022, 2024_Johnson, 2025_KAKOURI, MacBride_2025_CarForest, Hu2026xgboostinla}. Second, \emph{ensembles} combine separate predictions from statistical and ML/DL models into a single output \citep[e.g.,][]{Di2019, 2011GHEYAS}. 
Third, \emph{hierarchical integrations} embed both components within one probabilistic model; for instance, BAYESNF \citep{Saad2024} couples a neural network with hierarchical Bayesian inference by placing prior distributions on the network parameters.

The aim of this paper is to improve the predictive accuracy of spatio-temporal statistical models fitted via the INLA-SPDE approach, while preserving interpretability and uncertainty quantification. To this end, we propose two novel algorithms, collectively denoted as INLA-RF, which couple a parametric spatio-temporal LGM with a well-established ML algorithm as Random Forest (RF, \citealp{breiman2001random}). Both INLA-RF algorithms follow an iterative two-stage scheme: the INLA-SPDE based model captures spatio-temporal dependence, whereas RF learns remaining structure in the residuals and enhances predictive flexibility. Beyond the residual-learning backbone, we introduce explicit propagation of RF predictive uncertainty into the INLA-SPDE stage and propose two complementary feedback mechanisms: an offset-based mean correction (INLA-RF1) and a targeted node-level correction of the LGM latent field, focusing on stress points selected considering where the INLA-SPDE fit is most uncertain or least accurate (INLA-RF2).  In this regard, INLA-RF1 enhances predictive accuracy and uncertainty quantification when complex dependencies arise among explanatory variables. In contrast, INLA-RF2 refines the model structure via an extended RF-driven latent field, effectively capturing abrupt changes that the smooth components of a standard LGM may fail to accommodate.

For both INLA-RF algorithms we use the Kullback-Leibler divergence (KLD) as a stopping criterion. This means that the number of iterations of the iterative two-stage scheme is not a tuning parameter but is instead determined by an empirical measure of how much the posterior distribution changes between iterations. This approach can improve computational efficiency, by avoiding unnecessary iterations, and reduce the risk of overfitting.

We evaluate the predictive performance of the INLA-RF algorithms through two simulation studies and a real case study, using as benchmark non-hybrid models (i.e., an LGM fitted with INLA-SPDE and a standard RF). The results suggest that the INLA-RF algorithms can improve predictive performance, especially in the presence of temporal discontinuities or non-linear relationships between covariates and the response.

The paper is structured as follows. In Section~\ref{sec:PSTmodeling} we introduce the spatio-temporal framework we work in. Section~\ref{sec:backaground} provides the essential background and notation for INLA-SPDE and RF, limited to the elements required to follow our proposal; additional details are deferred to \ref{sec:reviewmodels} for the readers who are not familiar with these methods. Section~\ref{sec:INLA-RFalgos} presents the core contribution of the paper, namely the two novel INLA-RF algorithms (INLA-RF1 and INLA-RF2), together with their pseudo-code and implementation details. The empirical evaluation includes two simulation studies and a real case study with air pollution data, discussed in Section~\ref{sec:simulationstudy} and \ref{sec:realcasestudy}, respectively. Section~\ref{sec:discussion} concludes the paper with a discussion and final remarks. Supplementary material is provided in the appendices: \ref{AppendixB:metrics} presents the predictive performance metrics and the KLD-based stopping criterion, \ref{Sec:simulations_details} reports details about the simulation case studies, while 
\ref{Sec:realcase_details} contains supplementary results regarding the real case study.


\section{Methodology}\label{sec:Methdology}
\subsection{Spatio-temporal modeling framework}\label{sec:PSTmodeling}
We denote by $y(\mathbf s_i,t)$ the response variable at the $i^{th}$ location (with coordinates $\mathbf s_i \in \mathbb R^2$, $i = 1, \ldots, n$) and time $t \in \mathbb N$  ($t=1,\ldots, T$). Considering all the locations and time points, we obtain the following response vector  
\[
\mathbf y = \left(y(\mathbf s_1,1), \ldots, y(\mathbf s_n,1), \ldots, y(\mathbf s_1,T), \ldots, y(\mathbf s_n,T)\right)^{\top},
\]
 where $\top$ is the transpose operator. Moreover, $\mathbf{z}(\mathbf s_i,t) = \left(z_1(\mathbf s_i,t),\ldots, z_P(\mathbf s_i,t)\right)^\top$ represents the location-time specific $P$-dimensional covariate vector. 

We assume that the data are a function of the so-called large-scale component (LS), driven by covariates, and a random effects (RE) term that captures residual structure, including spatial and temporal dependence and measurement noise:
\begin{equation}\label{Eq:generalmodel}
y(\mathbf s_i,t)  =  LS(\mathbf s_i,t) + RE(\mathbf s_i,t).
\end{equation}
Even if $LS(\mathbf{s}_i,t)$ can be any (possibly non-linear) function of the covariates, a common approach is to assume that it is a linear function of the $P$ predictors, i.e., $LS(\mathbf s_i,t)=\boldsymbol{\beta}\mathbf z(\mathbf s_i,t)$, where $\boldsymbol{\beta}$ is the vector of fixed effect coefficients. 
The residual RE term is usually written as
$$RE(\mathbf{s}_i,t)=\omega(\mathbf{s}_i,t)+\epsilon(\mathbf{s}_i,t),$$
where $\omega(\mathbf{s}_i,t)$ is a spatio-temporal Gaussian process and $\epsilon(\mathbf{s}_i,t)$ is an unstructured (IID) measurement error. The spatio-temporal process in turn can be specified hierarchically in different ways \citep{STwithR2019}. In general, this leads, at each time point, to a zero-mean multivariate Gaussian distribution with a dense covariance matrix, with entries determined by the spatio-temporal covariance function. For this reason, the estimation of such a model can be computationally challenging, especially when dealing with a large number of spatial locations. A viable solution for this computational challenge, also known as \textit{big n problem} \citep{JonaLasinio2013}, is represented by the INLA-SPDE approach (see \ref{sec:inlaspde} for more details). It has been proved \citep[e.g.,][]{lindgren_spde, Lindgren:spde2022} as a valid and computationally efficient solution for fast LGM estimation and spatial prediction. In particular, since the INLA-SPDE approach adopts a Bayesian framework, it provides posterior distributions for all unknown parameters, including the spatially and temporally structured random effects, with uncertainty quantification being straightforward. Moreover, it allows for posterior predictive distributions of the response variable, enabling spatial prediction at any location within the domain of interest. 


The INLA-SPDE approach is already a powerful and computationally efficient solution for spatio-temporal modeling, providing fast Bayesian inference and principled uncertainty quantification. However, even when the spatio-temporal dependence is adequately captured, prediction accuracy can still be limited by the functional form chosen for the large-scale component. To push performance beyond this well-established framework, we complement it with RF, a nonparametric method that can capture complex nonlinearities and interactions left in the residuals. By feeding the RF-based residual correction back into the Bayesian model, our hybrid strategy delivers improved predictions without sacrificing the strengths of INLA-SPDE in terms of interpretability and uncertainty quantification.

\subsection{Essential background and notation about INLA-SPDE and RF}\label{sec:backaground}
As mentioned before, the proposed hybrid algorithms rely on two well-established methodologies: the INLA-SPDE approach (here after simply referred to as INLA) and RF. In this section we introduce the essential background and notation required to follow our proposal. Additional details are provided in \ref{sec:reviewmodels} for the readers who are not familiar with these methods.

We denote by $\mathrm{LGM}(\mathbf{A},\mathbf{x},\boldsymbol{\theta})$ the latent Gaussian model fitted with INLA, where $\mathbf{x}$ is the latent field which collects all the model latent components (fixed and random effects from the LS and RE terms of Eq.~\eqref{Eq:generalmodel}), $\boldsymbol{\theta}$ is the vector of hyperparameters, and $\mathbf{A}$ is the projector matrix linking observations to latent nodes (i.e., to the components of the latent vector $\mathbf{x}$). The linear predictor is thus given by $\boldsymbol{\eta}=\mathbf{A}\mathbf{x}$. INLA provides a Gaussian approximation of the posterior distribution of the latent field, which we summarize through its mean $\boldsymbol{\mu}$ and precision matrix $\mathbf{Q}$. These quantities are used to assess convergence of the INLA-RF algorithms via the Kullback-Leibler divergence (see \ref{sec:KLD}).

RF is an ensemble of regression trees; each tree is fit on a bootstrap resample of the training data \citep{breiman2001random}. On average, about one third of the training observations are left out of a given resample; these are referred to as out-of-bag (OOB) observations for each tree. By predicting each training point using only the trees for which it was OOB, it is possible to obtain out-of-sample predictions without an explicit data splitting (training vs test); the corresponding OOB mean squared error (OOB-MSE) is used here as an estimate of the RF prediction uncertainty.

\subsection{INLA-RF hybrid algorithms}\label{sec:INLA-RFalgos}

In this section we introduce the two hybrid modeling strategies we propose, collectively referred to as INLA-RF algorithms. The key idea is to alternate between two steps: fitting the parametric spatio-temporal LGM with INLA (STEP~1), and training RF on the resulting residuals to learn remaining structure (STEP~2). The RF output is then fed back into the INLA-based model, yielding an iterative residual-learning scheme. 
The two steps are repeated until the stopping condition $D_{KL}<\delta$ is met, where $D_{KL}$ is based on the Kullback-Leibler divergence \citep{Kullback_InformationTheory_1997}, and $\delta$ is a tolerance threshold (see \ref{sec:KLD} for the details about the computation of $D_{KL}$). If $D_{KL}$ is lower than $\delta$, it suggests that there is no significant difference in the posterior distribution of the latent field between two consecutive iterations, and further correction is not needed. 

The two algorithms are detailed in Section~\ref{Sec:INLARF1} and \ref{Sec:INLARF2}. For simplicity, in what follows we omit spatial and temporal indices and write $y(\mathbf{s}_i,t)$ as $y$; boldface denotes vectors and matrices.


\subsubsection{INLA-RF1: offset approach}\label{Sec:INLARF1}

INLA-RF1 takes inspiration from the SPAR-Forest approach proposed by \cite{MacBride_2025_CarForest} for the case of areal data, while here we focus on point-referenced data. At each iteration, RF is used to estimate a correction term that is incorporated into the INLA-based model as an offset.


More precisely, after initialization each iteration alternates the following two steps (see Algorithm~\ref{alg:SPDE_RF_1} for the corresponding pseudo-code):
\begin{enumerate}

\item \textbf{STEP 1} (INLA with offset): given the RF residual predictions from the previous iteration, $\hat{\mathbf e}^{(i-1)}_{\mathrm{RF}}$, we fit the INLA-based model by incorporating this term as an offset, i.e., a known constant that shifts the linear predictor and operates like a correction of the mean. Equivalently, at the $i^{th}$ iteration we compute the updated response
$
\mathbf y^{(i)}=\mathbf y -\hat{\mathbf e}^{(i-1)}_{\mathrm{RF}}
$
and fit the $\mathrm{LGM}(\mathbf A,\mathbf x,\boldsymbol\theta)$ with INLA using $\mathbf y^{(i)}$ as response and with the covariates entering linearly in the linear predictor. From the INLA output we retain the Gaussian approximation of the posterior distribution of the latent field $\mathbf x$, summarized by its mean $\boldsymbol\mu^{(i)}$ and precision matrix $\mathbf Q^{(i)}$, which are used to evaluate the KLD-based stopping criterion using $D_{KL}$.

We also obtain the fitted values $\hat{\mathbf y}^{(i)}_{\mathrm{INLA}}$ (given by the posterior mean of the predictive distribution) and compute the residuals to be passed to STEP~2, i.e. 
$
\hat{\mathbf e}^{(i)}_{\mathrm{INLA}}
= \mathbf y -\hat{\mathbf y}^{(i)}_{\mathrm{INLA}}.
$

\item \textbf{STEP 2} (RF on residuals): we train RF using the INLA residuals ${\hat{\mathbf e}_\mathrm{INLA}}^{(i)}$ as response and the covariates used in STEP~1 (augmented with spatial coordinates and time index) as predictors. This yields the predicted residuals $\hat{\mathbf e}^{(i)}_{\mathrm{RF}}$, which are fed back into STEP~1 at the next iteration.
\end{enumerate}


The algorithm is initialized as follows: $D_{KL}=10$, $\delta = 0.01$. Moreover, the initial values for $\boldsymbol\mu^{(0)}$, $\mathbf Q^{(0)}$ are obtained from an initial run of the LGM model fitted with $\mathbf{y}$ as response and including all the available predictors. 
The final hybrid predictor is obtained by adding the RF residual correction from STEP~2 to the INLA fitted values from STEP~1, i.e.,
$\hat{\mathbf y}=\hat{\mathbf y}_{\mathrm{INLA}}+\hat{\mathbf e}_{\mathrm{RF}}$.

An important issue when combining the two models concerns uncertainty propagation. By default, the RF output used in STEP~1 is incorporated into the INLA-based model as a fixed offset correction and therefore does not account for the uncertainty of the RF predictions. To incorporate this additional source of variability, when uncertainty propagation is enabled, during the INLA fit we use the INLA \texttt{precoffset} option\footnote{See \url{https://inla.r-inla-download.org/r-inla.org/doc/likelihood/gaussian.pdf}.} to include a fixed variance component in the Gaussian likelihood, taking the OOB-MSE $\hat{\sigma}^{2(i-1)}_{\mathrm{RF}}$ from the previous RF fit as a plug-in variance for the RF correction. This means that more uncertain RF corrections are accompanied by a larger observation-level variance contribution, effectively reducing the influence of the offset correction. 

\begin{algorithm}[h!]
\caption{INLA-RF1: offset approach}
\label{alg:SPDE_RF_1}
\hypertarget{alg:SPDE_RF1}{}

\textbf{State 1:} Set $D_{KL}=10$ and  $\delta=0.01$. Fit with INLA an initial LGM($\mathbf A,\mathbf x,\boldsymbol\theta$) using $\mathbf y$ as response; obtain
$\boldsymbol\mu^{(0)},\mathbf Q^{(0)}$ and the fitted values $\hat{\mathbf y}^{(0)}_{\mathrm{INLA}}$.\\
\textbf{State 2:} Compute residuals
$\hat{\mathbf e}^{(0)}_{\mathrm{INLA}}=\mathbf y-\hat{\mathbf y}^{(0)}_{\mathrm{INLA}}$ and train RF on $\hat{\mathbf e}^{(0)}_{\mathrm{INLA}}$
to obtain $\hat{\mathbf e}^{(0)}_{\mathrm{RF}}$ and the OOB-MSE $\hat{\sigma}^{2(0)}_{\mathrm{RF}}$.\\
\textbf{State 3:} Set $i=1$.\\

\While{$D_{KL} \ge \delta$}{
\vspace{5pt}
\textbf{STEP 1} (INLA with offset): \\
Compute the updated response
$\mathbf y^{(i)}=\mathbf y-\hat{\mathbf e}^{(i-1)}_{\mathrm{RF}}$.\\

\If{uncertainty propagation == TRUE}{
Fit the LGM($\mathbf A,\mathbf x,\boldsymbol\theta$) using $\mathbf y^{(i)}$ as response and including a fixed variance component given by $\hat{\sigma}^{2(i-1)}_{\mathrm{RF}}$ (OOB-MSE) in the Gaussian likelihood through \texttt{precoffset}.
}
\Else{
Fit the LGM($\mathbf A,\mathbf x,\boldsymbol\theta$) using $\mathbf y^{(i)}$ as response.
}
        \vspace{5pt}

    Obtain $\boldsymbol\mu^{(i)},\mathbf Q^{(i)}$ and compute the KLD-based convergence metric $D_{KL}$.        \\
    

    Obtain the fitted values $\hat{\mathbf y}^{(i)}_{\mathrm{INLA}}$ for the updated response and compute the INLA residuals
    $\hat{\mathbf e}^{(i)}_{\mathrm{INLA}}=\mathbf y-\hat{\mathbf y}^{(i)}_{\mathrm{INLA}}$.\\
    
    \textbf{STEP 2} (RF on residuals): 
    Train RF using $\hat{\mathbf e}^{(i)}_{\mathrm{INLA}}$ as response and obtain the predictions $\hat{\mathbf e}^{(i)}_{\mathrm{RF}}$
    and the OOB-MSE $\hat{\sigma}^{2(i)}_{\mathrm{RF}}$.\\
    
    $i=i+1$.
}
\vspace{5pt}
\textbf{Output:} the final hybrid predictor is obtained by adding the RF residual correction from STEP~2 to the INLA fitted values from STEP~1, i.e.,
$\hat{\mathbf y}=\hat{\mathbf y}_{\mathrm{INLA}}+\hat{\mathbf e}_{\mathrm{RF}}$.

\end{algorithm}

\subsubsection{INLA-RF2: latent field nodes correction}\label{Sec:INLARF2}
INLA-RF2 uses the RF output to correct a selected subset of latent field nodes. 
Unlike INLA-RF1, which applies a mean correction through an offset in the linear predictor, INLA-RF2 performs a \emph{localized} adjustment by augmenting the LGM with an additional IID random effect indexed only on a selected set of indices $\mathcal K$ (hereafter also referred to as \textit{locations} or \emph{stress nodes}), corresponding to the nodes identified for the correction. Identifying the stress nodes is the most delicate step of the algorithm. To the authors’ knowledge, there is no single, well-defined criterion for selecting either the number or the location of these nodes. The difficulty arises from the fact that observations are linked to latent nodes through the projector matrix, so that the correspondence between observations and nodes is not one-to-one. Moreover, the number of observations associated with each node may vary, and different components of the latent field may be candidates for correction.

In the INLA-RF2 construction, the RF-based information enters the model through an explicit latent component, thereby enabling a straightforward uncertainty propagation within the INLA framework. More details on latent field nodes and on the construction of the required projector matrices are provided in \ref{sec:inlaspde}.

The algorithm starts from an initial INLA fit of a LGM($\mathbf{A}, \mathbf{x}, \boldsymbol{\theta}$) using $\mathbf y$ as response); this is used both to identify the latent field nodes to be corrected and to initialize the posterior mean and precision quantities ($\boldsymbol\mu^{(0)}$ and $\mathbf Q^{(0)}$) required by the stopping criterion.
%
Let $\mathbf x_c=\{x_k: k\in\mathcal K\}$ denote the subset of latent components targeted for correction, where the cardinality $|\mathcal K|$ is fixed a priori (e.g., $|\mathcal K|=100$). Nodes can be selected according to different criteria, such as the largest marginal posterior variances $Var(x_k\mid\mathbf y)$ or the largest predictive errors associated with the corresponding part of the linear predictor. 
%
%
For each selected node, an additional correction term is introduced, yielding a new random effect denoted by $\mathbf{x}'_c$. This additional component is indexed by the selected nodes in $\mathcal K$, but is represented in the model as a separate random effect. In particular, an IID Gaussian prior distribution is assumed for the correction component $\mathbf{x}'_c$: 
\begin{equation}\label{eq:parametersxcprime}
\mathbf{x}'_c \sim \text{MVN}\left(\boldsymbol{\mu}'_c, (\tau^\prime_c)^{-1}\mathbf{I}\right).
\end{equation}
In particular, at the $i^{th}$ iteration of the algorithm we have that
\begin{equation}
    \begin{split}
        {\boldsymbol\mu_c^\prime}^{(i)} & = {\boldsymbol\mu_c^\prime}^{(i-1)} + \hat{\mathbf{e}}^{(i-1)}_\mathrm{RF}\\
{\tau^\prime}^{(i)}_c &= \left(\hat{\sigma}_{\mathrm{RF}}^{2(i-1)}\right)^{-1},
    \end{split}
\end{equation}
%
%
%
where $\hat{\sigma}_{\mathrm{RF}}^{2(i-1)}$ is the RF OOB-MSE and $\hat{\mathbf{e}}^{(i-1)}_{RF}$ denotes the RF node-level correction vector obtained as follows. First, RF predicts the INLA residuals at the observation level. Then, each predicted residual is equally split among the selected nodes linked to that observation, and the resulting contributions are averaged over the observations associated with each selected node. From a practical point of view, in INLA each random effect is specified with zero mean. Therefore, in the actual implementation the formulation in Eq.~\eqref{eq:parametersxcprime} is represented equivalently by combining a zero-mean random effect with an offset term carrying the node-specific correction vector $\boldsymbol\mu_c^\prime$, which is zero everywhere except for the entries associated with the added correction component.


After initialization, each iteration alternates the following two steps (see Algorithm~\ref{alg:SPDE_RF_2} for the corresponding pseudo-code):
\begin{enumerate}
    \item in \textbf{STEP~1} we use the RF residual predictions $\hat{\mathbf{e}}^{(i-1)}_\mathrm{RF}$ to update the node-level correction vector  $\boldsymbol{\mu}_c^{\prime(i)}$ and the associated precision parameter $\tau_c^{\prime(i)}$ (see Eq.~\eqref{eq:parametersxcprime}). The correction component ${\mathbf{x}'_c}^{(i)}$ is specified as an additional IID random effect in the enlarged LGM ($\{\mathbf{A},\mathbf{A}_c\}, \{\mathbf{x}, {\mathbf{x}_c^\prime}^{(i)}\}, \boldsymbol{\theta}$), fitted with INLA using $\mathbf{y}$ as response. The matrix
$\mathbf{A}_c$ maps the added IID correction effect to the observations linked to the selected nodes
    
    As in Algorithm~\ref{alg:SPDE_RF_1}, we conclude the first step by computing the INLA predictions and residuals ${\hat{\mathbf e}_\mathrm{INLA}}^{(i)} = \mathbf y - {\hat{\mathbf y}_\mathrm{INLA}}^{(i)}$, and by obtaining $\boldsymbol\mu^{(i)}$ and $\mathbf Q^{(i)}$ which are needed for computing $D_{KL}$ and evaluating the stopping criterion.
    \item In \textbf{STEP~2} we train an RF using the INLA residuals ${\hat{\mathbf e}_\mathrm{INLA}}^{(i)}$ as response variable and the time index together with the spatial coordinates as predictors. This yields observation-level residual predictions, which are then transferred to the selected nodes by equally splitting each predicted residual among the linked nodes and averaging the resulting contributions for each node.
\end{enumerate}

The final output of Algorithm~\ref{alg:SPDE_RF_2} is the last enlarged LGM fit with INLA, where the RF based correction has already been incorporated.

\begin{algorithm}[h!]
\caption{INLA-RF2: latent field nodes correction} \label{alg:SPDE_RF_2}
\hypertarget{alg:SPDE_RF2}{}
\textbf{State 1:} Set $D_{KL} = 10$, and $\delta = 0.01$. Fit with INLA an initial LGM($\mathbf{A}, \mathbf{x}, \boldsymbol{\theta}$) using $\mathbf{y}$ as response; obtain $\boldsymbol\mu^{(0)}$, $\mathbf Q^{(0)}$ and the fitted values ${\hat{\mathbf y}^{(0)}_\mathrm{INLA}}$.\\
\textbf{State 2:} Set $|\mathcal{K}|$ and determine the nodes to be corrected $\mathbf{x}_c=\{x_k: k \in \mathcal{K}\}$.\\
\textbf{State 3:} Compute the INLA residuals
${\hat{\mathbf e}^{(0)}_\mathrm{INLA}} = \mathbf y - {\hat{\mathbf y}^{(0)}_\mathrm{INLA}}$
and train RF to obtain ${\hat{\mathbf e}^{(0)}_\mathrm{RF}}$ and $\hat{\sigma}^{2(0)}_\mathrm{RF}$.\\
\textbf{State 4:} Set $i = 1$.\\

\vspace{5pt}
\While{$D_{KL} \geq \delta$}{\vspace{1mm}

\textbf{STEP 1}: use the RF output to update the node-level correction and its associated precision:
${\boldsymbol\mu_c^\prime}^{(i)}  = {\boldsymbol\mu_c^\prime}^{(i-1)} + \hat{\mathbf{e}}^{(i-1)}_\mathrm{RF}$ and
${\tau^\prime}^{(i)}_c = \left(\hat{\sigma}_{\mathrm{RF}}^{2(i-1)}\right)^{-1}$. Define the additional latent component as
${\mathbf{x}_c^\prime}^{(i)} \sim \text{MVN}\left(\boldsymbol{\mu}_c^{\prime(i)}, (\tau'_c)^{-1}\mathbf{I}\right)$.\\

Fit the enlarged LGM ($\{\mathbf{A},\mathbf{A}_c\}, \{\mathbf{x}, {\mathbf{x}_c^\prime}^{(i)}\}, \boldsymbol{\theta}$) using $\mathbf{y}$ as response.\\

 Obtain $\boldsymbol\mu^{(i)},\mathbf Q^{(i)}$ and compute the KLD-based convergence metric $D_{KL}$.        \\

    Obtain the fitted values $\hat{\mathbf y}^{(i)}_{\mathrm{INLA}}$ and compute the INLA residuals
    $\hat{\mathbf e}^{(i)}_{\mathrm{INLA}}=\mathbf y-\hat{\mathbf y}^{(i)}_{\mathrm{INLA}}$.\\

\textbf{STEP 2}: implement RF with ${\hat{\mathbf e}_\mathrm{INLA}}^{(i)}$ as response and obtain observation-level residual predictions; aggregate these predictions over the observations linked to the selected nodes to obtain ${\hat{\mathbf e}_\mathrm{RF}}^{(i)}$.\\

$i=i+1$
}
\textbf{Output:} the final predictor is given by the fitted values of the last enlarged INLA-based model, where the RF-based correction has already been incorporated, i.e., $\hat{\mathbf y} = \hat{\mathbf y}_\mathrm{INLA}$.
\end{algorithm}

\section{Simulation studies}\label{sec:simulationstudy}

To evaluate the performance of the INLA-RF algorithms, we conduct two simulation studies. The first is a spatio-temporal example which compares the INLA-RF1 algorithm with a spatio-temporal LGM with no hybridization and with the standard RF algorithm. The second is a purely temporal scenario designed to assess the INLA-RF2 approach in modeling temporal discontinuities. In this simulation study, the comparison focuses on the benchmark INLA-based
model only, since the aim is to assess the ability of INLA-RF2 to improve the posterior inference of a given Bayesian latent field through a targeted local correction. In this setting, a standard RF model is not a directly comparable
benchmark, as it does not operate on the same latent component nor provide inference on the same posterior quantities.

For the comparisons we use the following performance metrics: Root Mean Square Error (RMSE), Mean Absolute Error (MAE), Coverage Probability (CP), and Average Interval Width (AIW); see \ref{AppendixB:metrics} for their definitions. In the implementation of Algorithm~\ref{alg:SPDE_RF_1}, both with and without uncertainty propagation, and Algorithm~\ref{alg:SPDE_RF_2} we use the KLD stopping criterion as defined in Eq.~\eqref{eq:averageKLD}.

\subsection{Spatio-temporal simulation study for INLA-RF1} \label{Sec:sim1}

\begin{figure}[h!]
    \centering
    \includegraphics[scale=0.35]{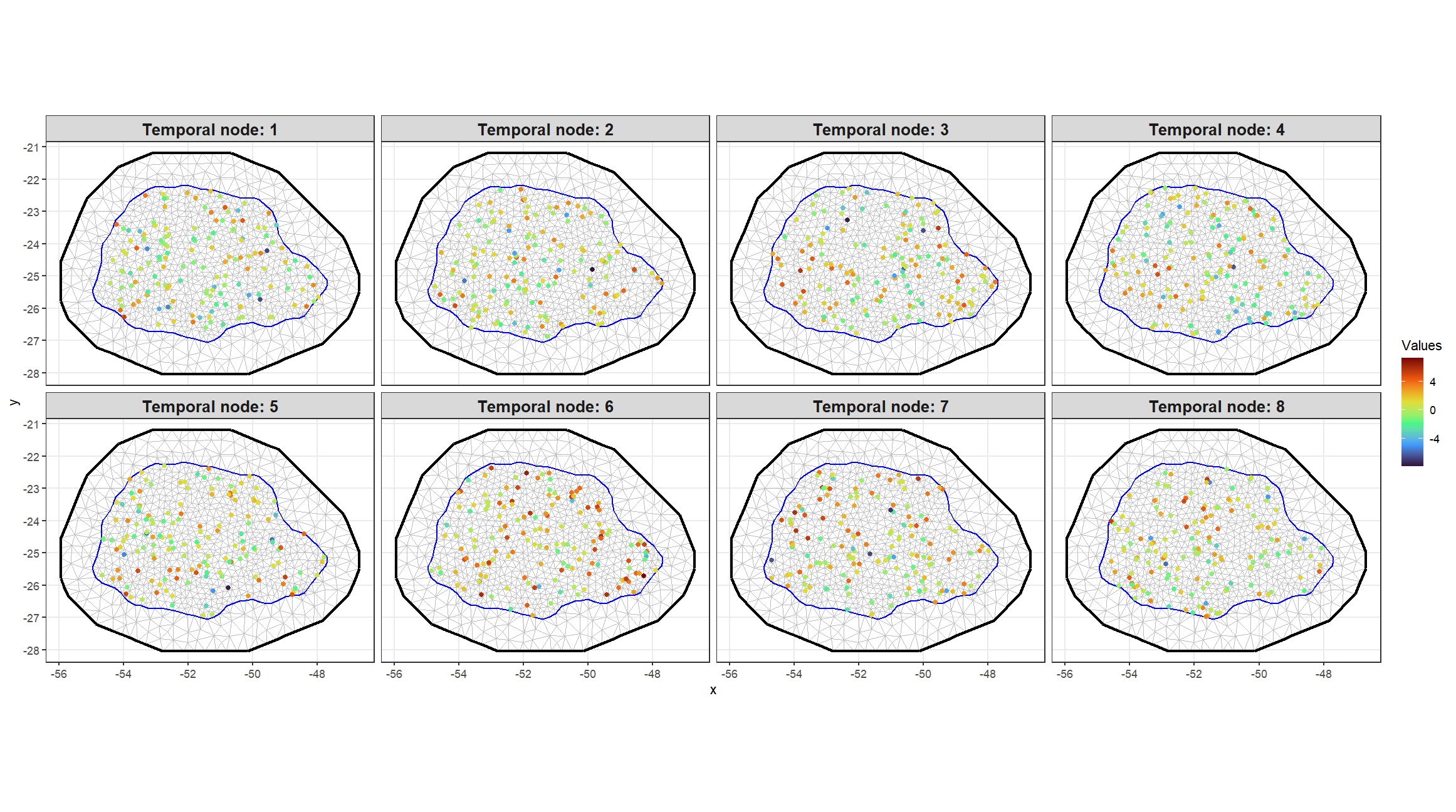}
    \caption{Simulated data for the spatio-temporal simulation study for INLA-RF1 with $n=150$ and $T=8$ (locations changing randomly in time), together with the SPDE mesh used for INLA fitting.}
    \label{fig:simulateddata}
\end{figure}

In this case study we simulate a dataset with $T=8$ time points and $n=150$ spatial locations at each time location. The response is assumed to be normally distributed, with the linear predictor being a non-linear function of 3 predictors. The data-generating mechanism and simulation settings are described in \ref{Appendix:sim1data}. The simulated data for the eight time points are shown in Figure~\ref{fig:simulateddata}, together with the SPDE mesh used for INLA fitting. $80\%$ of the simulated data are used for training while the rest for testing.

As a benchmark, and in STEP 1 of the INLA-RF1 algorithm, we consider a parametric LGM analogous to the spatio-temporal model used to simulate the data (see Eq.~\eqref{Eq:simulationmodel}), but assuming that the three covariates enter linearly in the linear predictor.

The prior distributions for Bayesian inference were chosen to be vague for the fixed effects, using Gaussian priors with high variance, i.e. $\text N(0, \sigma^2 = 1/0.001)$. For the spatio-temporal effects, we have three hyperparameters: $\sigma$, $\rho$ and $a$. For $\sigma$ and $\rho$, we use Penalized Complexity (PC) priors \citep{Simpson_2017_PCprior, Fuglstad_2019_PCGF}, specified as $P(\sigma > \sigma_0) = 0.5$ and $P(\rho < \rho_0) = 0.5$, where $\sigma_0 = 1$ and $\rho_0$ is set to the maximum distance ($d$) between two points on the boundary of the study region divided by 5, i.e., $\rho_0 = d/5$. Finally, the prior distribution for the temporal autocorrelation parameter ($a$) is defined as $\text{logit}(a) \sim \text{N}(0, 0.15)$.

\subsubsection{Results}



Table~\ref{tab:TrainTest_Measures_Performance} presents the results obtained for the 80\%-20\% training-test split (the INLA label is used for the benchmark INLA-based model with no hybridization, while INLA-RF1.1 and INLA-RF1.2 refer to the INLA-RF1 algorithm without and with uncertainty propagation, respectively). From this table, we can observe that applying INLA-RF1, regardless of uncertainty propagation, improves the RMSE, MAE, and CP metrics, both for training and test data. In particular, the AIW metric suggests that when the hybrid approach is applied without uncertainty propagation (INLA-RF1.1), the predictions are more precise; this is because the mean estimates are more accurate, although the uncertainty from the RF is not propagated to the INLA output. However, when uncertainty is propagated (INLA-RF1.2), the significant improvement in CP is accompanied by a wider predictive distribution, as indicated by a larger AIW, compared to the standard approach without hybrid inference. 


\begin{table}[ht]
\centering
\begin{tabular}{lcccccccc}
\toprule
& \multicolumn{2}{c}{RMSE} & \multicolumn{2}{c}{MAE} & \multicolumn{2}{c}{CP} & \multicolumn{2}{c}{AIW} \\
\cmidrule(lr){2-3} \cmidrule(lr){4-5} \cmidrule(lr){6-7} \cmidrule(lr){8-9}
\textbf{Model} & \textbf{Train} & \textbf{Test} & \textbf{Train} & \textbf{Test} & \textbf{Train} & \textbf{Test} & \textbf{Train} & \textbf{Test} \\
\midrule
INLA        & 1.28 & 1.80 & 0.97 & 1.30 & 0.68 & 0.59 & 2.19 & 2.45 \\
RF          & 1.06 & 1.11 & 0.82 & 0.86 & 0.95 & 0.95 & 4.38 & 3.97 \\
INLA-RF1.1  & 0.59 & 1.00 & 0.45 & 0.75 & 0.84 & 0.68 & 1.49 & 1.84 \\
INLA-RF1.2  & 0.60 & 1.01 & 0.46 & 0.76 & 0.97 & 0.88 & 2.78 & 2.99 \\
\bottomrule
\end{tabular}
\caption{Performance metrics based on an 80\%-20\% training-test split. The INLA label corresponds to the benchmark INLA-based model with no hybridization, while RF denotes the standard Random Forest. INLA-RF1.1 denotes the INLA-RF1 model without uncertainty propagation, whereas INLA-RF1.2 includes uncertainty propagation.}
\label{tab:TrainTest_Measures_Performance}
\end{table}

\begin{figure}[h!]
    \centering
    \includegraphics[scale=0.7]{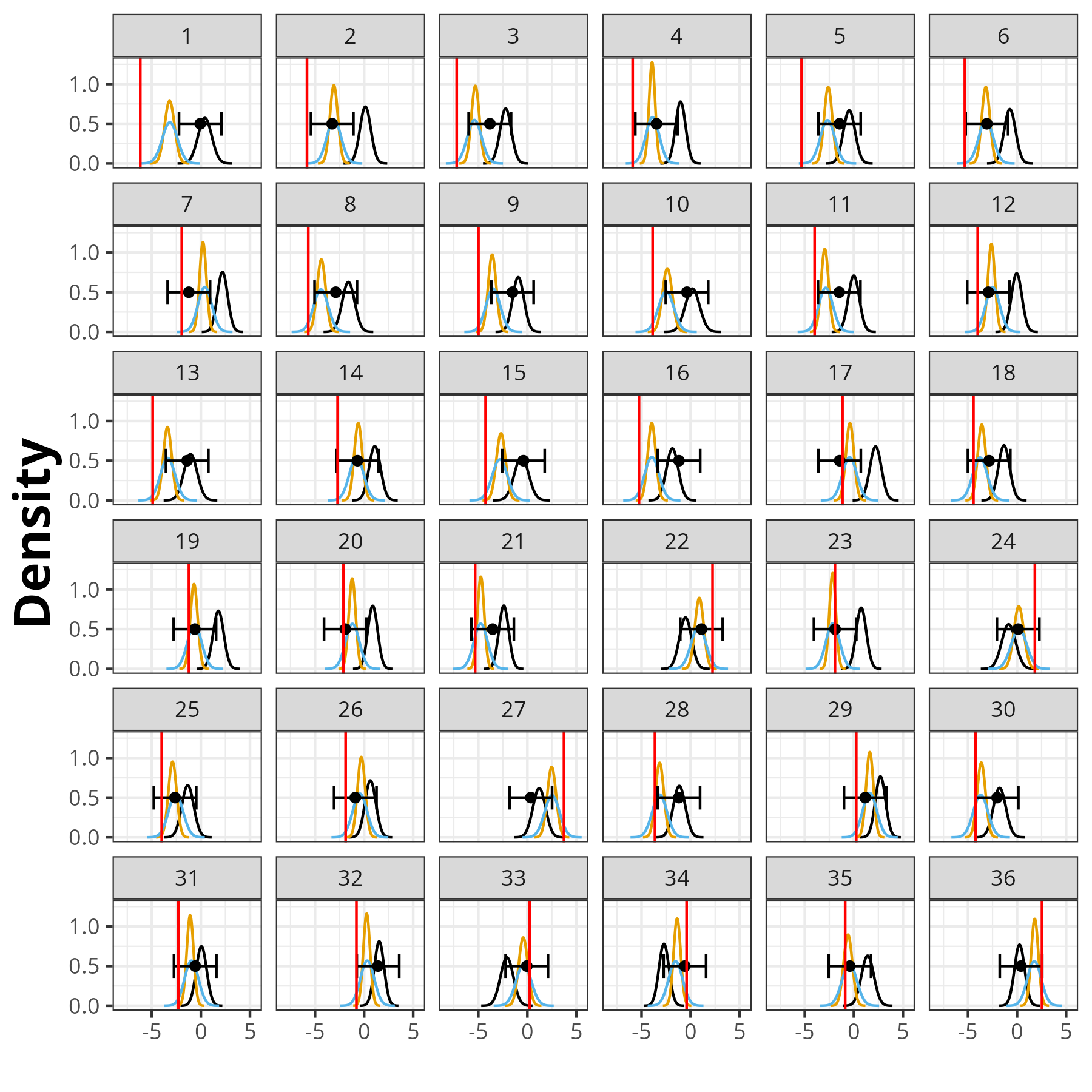}
    \caption{Posterior distributions of the linear predictor for the $36$ observations in the training data set with the highest RMSE values computed using the benchmark INLA-based model. The black solid line shows the benchmark INLA-based output, orange solid and light blue solid lines correspond to the output from Algorithm~\ref{alg:SPDE_RF_1} without and with uncertainty propagation (INLA-RF1.1 and INLA-RF1.2), respectively. Red vertical lines indicate the simulated (true) values and the horizontal black error bars correspond to the CI at $95\%$ of the RF output.}
    \label{fig:posterior_distribution_with_no_uncertainty}
\end{figure}

\begin{figure}[h!]
    \centering
    \includegraphics[scale=0.7]{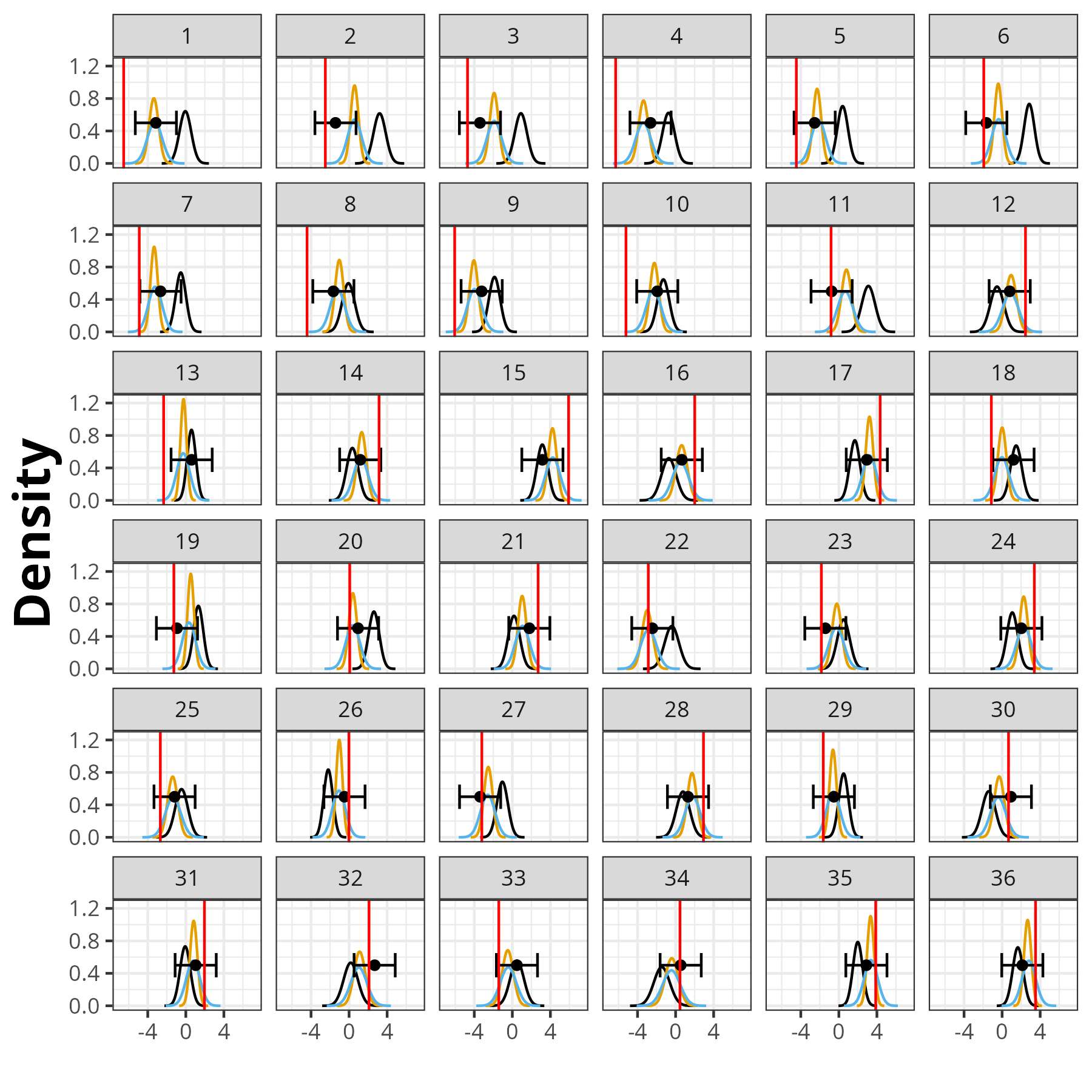}
 \caption{Posterior distributions of the linear predictor for the $36$ observations in the test data set with highest RMSE values computed using the benchmark INLA-based model. The black solid line shows the benchmark INLA-based output, orange solid and light blue solid lines correspond to the output from Algorithm~\ref{alg:SPDE_RF_1} without and with uncertainty propagation (INLA-RF1.1 and INLA-RF1.2), respectively. Red vertical lines indicate the simulated (true) values and the horizontal black error bars correspond to the CI at $95\%$ of the RF output.}
    \label{fig:posterior_distribution_with_uncertainty}
\end{figure}

Figures~\ref{fig:posterior_distribution_with_no_uncertainty} and~\ref{fig:posterior_distribution_with_uncertainty} show $36$ posterior predictive distributions for the observations with the highest RMSE under the benchmark INLA-based model (represented by the black solid line). The first figure refers to the training set, while the second shows results for the test set. In general, these figures illustrate that the posterior distributions obtained using INLA-RF1 algorithm without uncertainty propagation (INLA-RF1.1, orange solid lines) yield better alignment with the true simulated values (corresponding to the vertical red lines), though with reduced uncertainty. In contrast, when the algorithm is applied with uncertainty propagation (INLA-RF1.2, light blue solid lines), we observe an improvement in the localization of the true values, but at the cost of increased uncertainty (i.e., higher variance) compared to the INLA-based model. Finally, the INLA-RF1 algorithms, both with and without uncertainty propagation, achieve a more accurate localization of the true values compared with the RF algorithm (black horizontal error bars), with substantially lower uncertainty.

In \ref{AppendixA:CV} we perform a cross-validation analysis to evaluate the models' performance under a structured spatio-temporal $k$-fold partitioning \citep{Roberts_2017_CV, Meyer_2018_SpTCV} in order to take into account the spatio-temporal structure of the data when splitting the data into training and test. The results show that the average performance, as well as the performance for each individual CV partition, is consistent with the results obtained when using a standard ($80\%-20\%$) training-test split.

\subsection{Purely temporal simulation study for INLA-RF2}\label{Sec:sim2}

\begin{figure}[ht]
\centering
\includegraphics[width=1\linewidth]{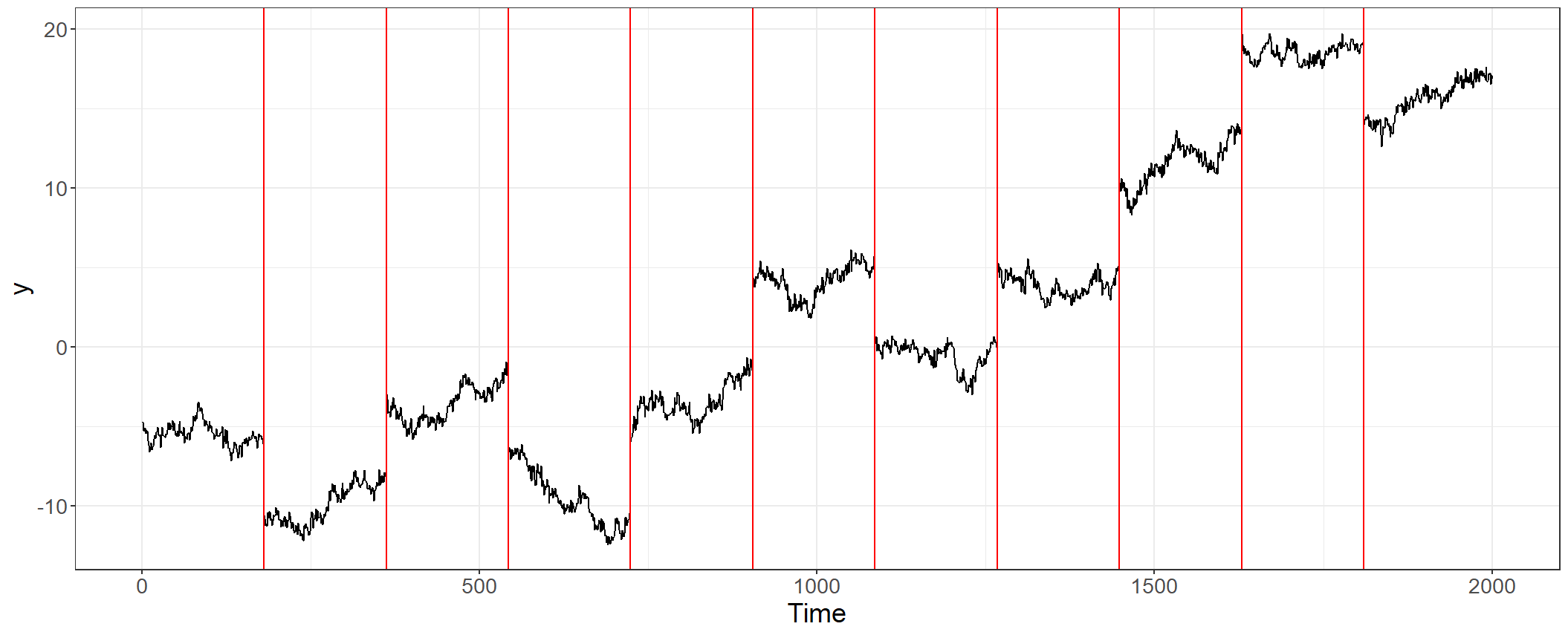}
\caption{Simulated time series with 2000 time points and 10 jumps (highlighted with red vertical lines).}
\label{fig:simulated_data}
\end{figure}

In this case study, we simulate a purely temporal dataset designed to provide a simple and controlled setting for illustrating the behavior of INLA-RF2 in the presence of abrupt temporal discontinuities, which may arise in practice because of policy interventions, regime changes, or external shocks. The data-generating mechanism and simulation settings are described in \ref{Appendix:sim2data}. The simulated time series with $n=2000$ time points and 10 jumps is reported in Figure~\ref{fig:simulated_data}.

The first step for implementing INLA-RF2 consists in fitting an initial LGM to identify the latent field nodes to be corrected (see States 1 and 2 in Algorithm~\ref{alg:SPDE_RF_2}); this model will also be used as benchmark for comparison. In this regard, we use the following purely temporal model:
\begin{equation}\label{eq:sim2_basemodel}
\mathbf{y} = \beta_0 \mathbf 1 + \mathbf{u} + \boldsymbol\varepsilon_\ell,
\end{equation}
where $\beta_0$ is the intercept and $\boldsymbol\varepsilon_\ell \sim \text{MVN}(\mathbf 0, \tau_{\ell}^{-1}\mathbf I)$. The term $\mathbf{u}$ is a second-order random walk process (RW2), meaning that its vector of second-order increments, denoted by $\Delta^2\mathbf{u}$, follows
$\Delta^2\mathbf{u} \sim \text{MVN}(\mathbf{0}, \tau_{u}^{-1}\mathbf{I})$,
where $(\Delta^2\mathbf{u})_t = u_t - 2u_{t-1} + u_{t-2}$. The prior distributions used to fit this model are the following: a vague Gaussian prior, $\mathrm{N}(0, \sigma^2 = 1/0.001)$, was specified for the intercept; for the precision $\tau_\ell$ and $\tau_u$ a $\mathrm{log\text{-}Gamma}(1, 5 \cdot 10^{-5})$ prior was assigned.



\begin{figure}[h]
\centering
\includegraphics[width=0.95\linewidth]{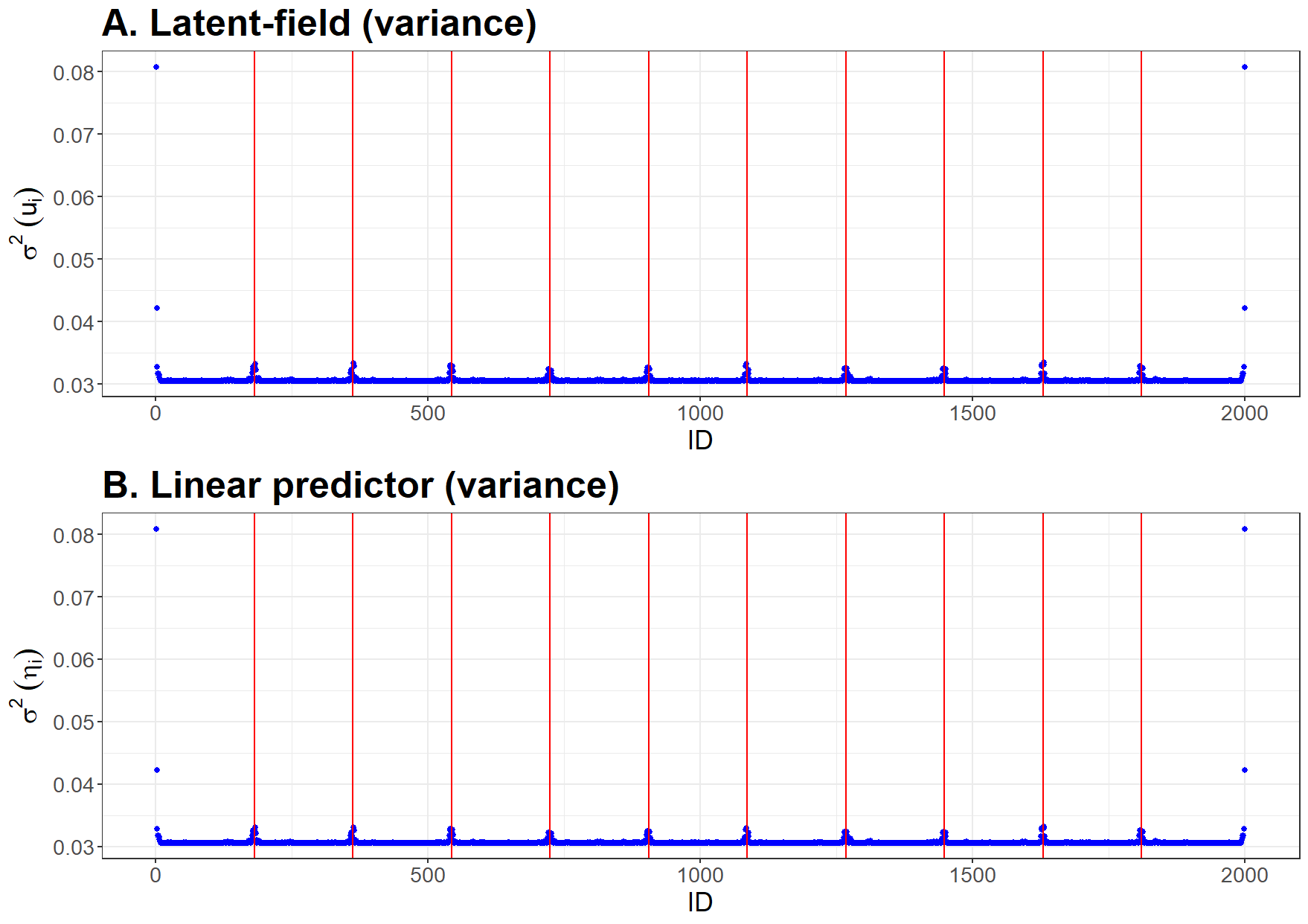}
\caption{Results from the initial INLA-based model (see Eq.~\eqref{eq:sim2_basemodel}): Panel A displays the marginal variance of the temporal random effect, while Panel B shows the marginal variance of the linear predictor. In all the panels the vertical red lines indicate the locations where jumps were introduced in the simulated data.}
\label{fig:model_base_latent_predictor}
\end{figure}

\begin{figure}
\centering
\includegraphics[width=1\linewidth]{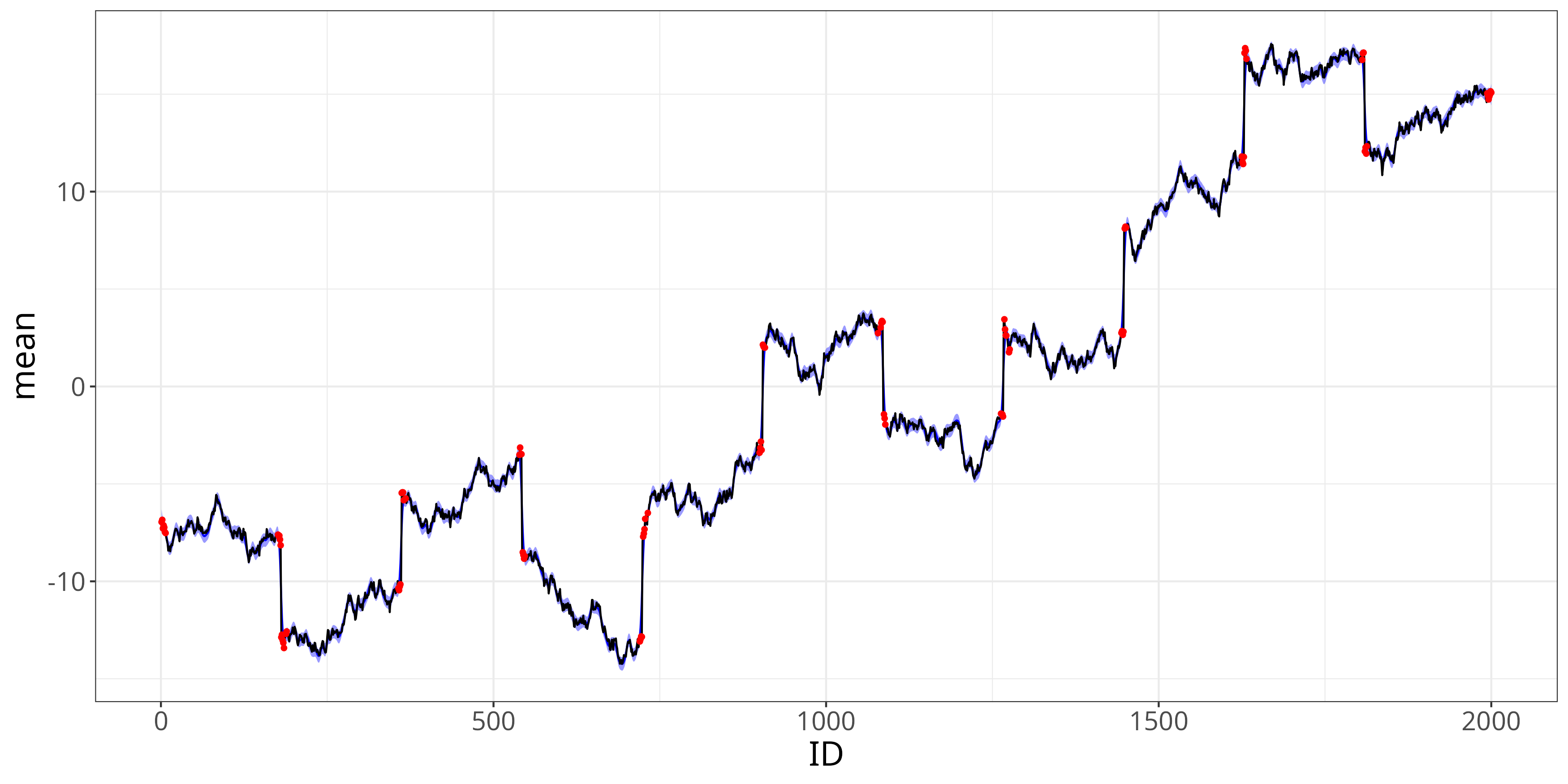}
\caption{Posterior mean of the temporal random effect estimated using the initial INLA-based model. The nodes selected as stress points according to the highest marginal variance criterion are highlighted in red.}
\label{fig:strees_points}
\end{figure}

The results obtained from this model for the latent field and the linear predictor $\boldsymbol\eta =  \beta_0 \mathbf 1 + \mathbf{u}$ are shown in Figure~\ref{fig:model_base_latent_predictor}. In particular, Panels A and B show the marginal variance values for each node of the temporal random effect ($\mathbf{u} = (u_1, \dots, u_{200})$) and of the linear predictor ($\boldsymbol\eta = (\eta_1, \dots, \eta_{200})$), respectively. The locations of the jumps are also indicated by vertical red lines. As expected, the largest marginal variances are observed at the boundary nodes, due to the structure of the precision matrix of the temporal component, and at nodes located close to the jumps. Based on the results in Panel A, we identify the \textit{stress points}, i.e., the latent field nodes to be corrected, by selecting the $|\mathcal{K}|=100$ nodes of the temporal random effect with the highest marginal variances, as shown in Figure~\ref{fig:strees_points}.

Having selected the nodes for correction, we proceed with the implementation of the INLA-RF2 algorithm. Following the general framework described in Section~\ref{Sec:INLARF2}, we extend the initial INLA-based model (Eq.~\eqref{eq:sim2_basemodel}) by introducing an additional correction component defined on the selected nodes. In this purely temporal setting, the enlarged model used in STEP 1 of Algorithm 2 is given by:
\begin{equation}
\boldsymbol y \sim \text{MVN}(\boldsymbol{\mu}, \tau_\ell^{-1} \mathbf{I}), \quad
\boldsymbol{\mu} = \beta_0 \mathbf{1} + \mathbf{u} + \mathbf{A}_c (\boldsymbol{\mu}'_c + \mathbf{x}'_c),
\end{equation}
where $\mathbf{A}_c$ maps the correction terms associated with the selected nodes to the corresponding observations. The correction vector $\boldsymbol{\mu}'_c$ is obtained from the RF predictions of the INLA residuals and is defined at the node level after aggregating observation-level predictions. The associated latent effect $\mathbf{x}'_c$ captures residual uncertainty around this correction, with precision determined by the RF OOB-MSE. The parameters are iteratively updated according to Algorithm~\ref{alg:SPDE_RF_2} until convergence.
Note that writing the correction term as $\boldsymbol{\mu}'_c + \mathbf{x}'_c$ is equivalent to specifying $\mathbf{x}'_c \sim \text{MVN}(\boldsymbol{\mu}'_c, (\tau'_c)^{-1}\mathbf{I})$, as described in Section~\ref{Sec:INLARF2}; however, in the INLA implementation the correction is decomposed into a deterministic component $\boldsymbol{\mu}'_c$ (introduced via an offset) and a zero-mean stochastic component $\mathbf{x}'_c$.




\subsubsection{Results}
\begin{figure}[h!]
\centering
\includegraphics[width=1\linewidth]{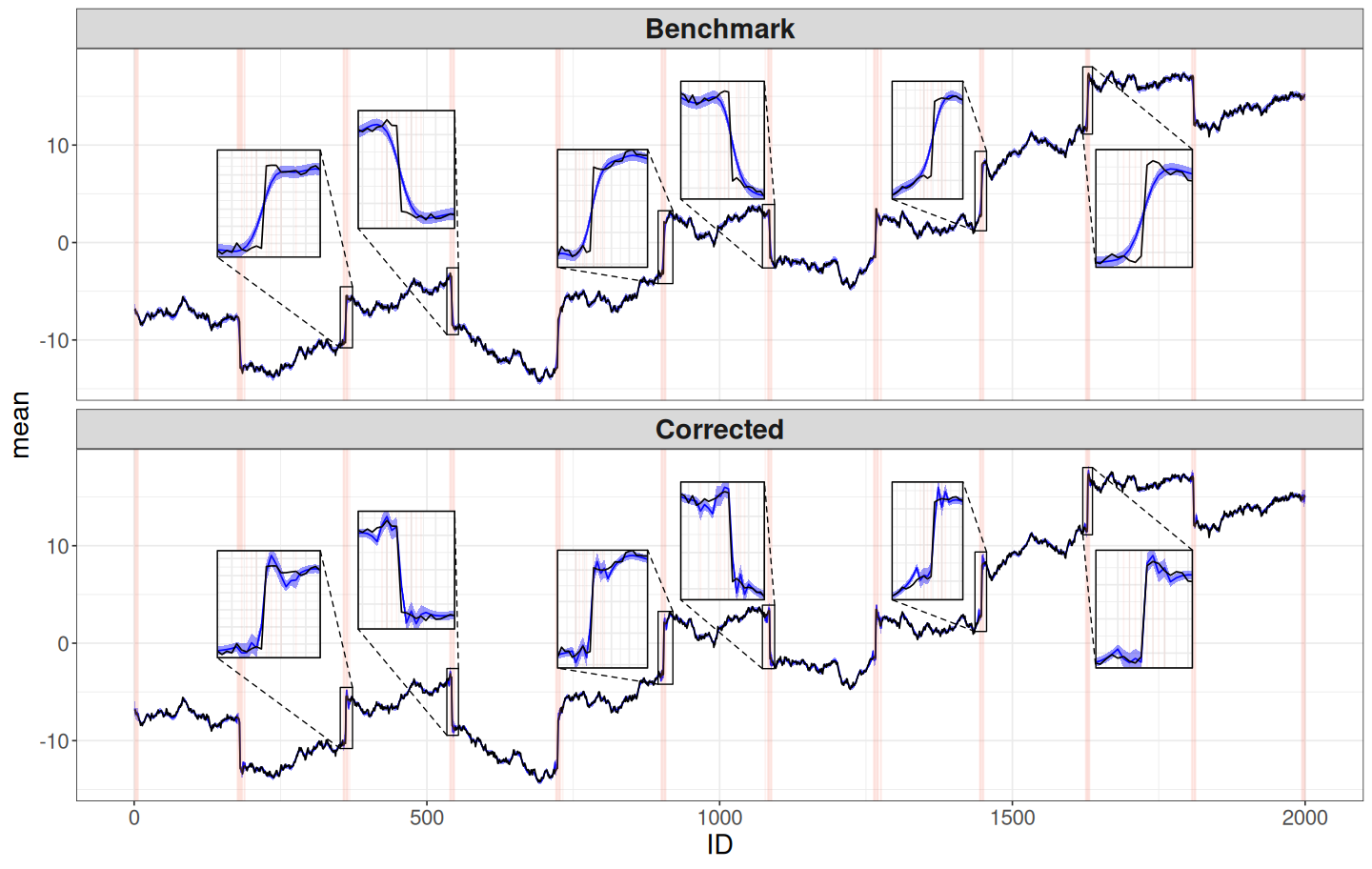}
\caption{Posterior mean of the temporal random effect under the INLA-RF2 model (bottom) and the benchmark INLA-based model (top), with $95\%$ credible intervals shown in blue and the simulated time series in black. The vertical red lines indicate the locations of the 100 stress points.}
\label{fig:corrected_latent_field_original}
\end{figure}

In this simulation study, the comparison focuses on the benchmark INLA-based model only.
Figure~\ref{fig:corrected_latent_field_original} shows the simulated time series (in black) together with the posterior mean and 95\% credible interval of the corrected temporal field obtained with INLA-RF2 and of the temporal field estimated using the benchmark INLA-based model. Zoomed-in views around selected jumps are also included to highlight local differences. In the regions near the imposed jumps, the corrected latent field provides a closer fit to the true temporal trend when compared with the estimates provided by the benchmark INLA-based model.

\begin{figure}[h!]
\centering
\includegraphics[scale=0.4]{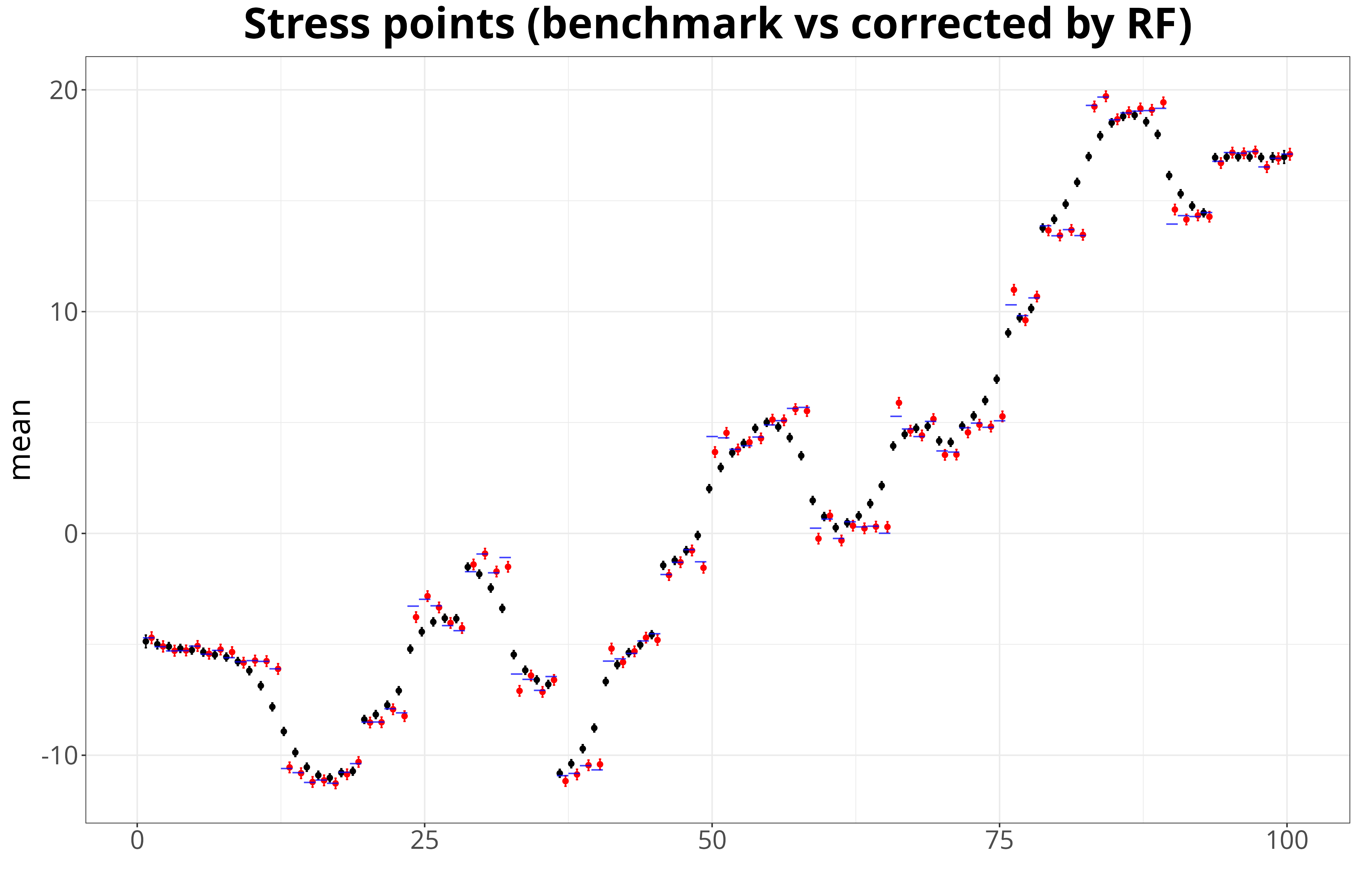}
\caption{Temporal field estimates at the 100 stress points under the benchmark INLA-based model and the INLA-RF2 model, compared with the true simulated values. Posterior means and 95\% credible intervals from the benchmark INLA-based model are shown in black, while those from the INLA-RF2 model are shown in red. The true simulated values are displayed as horizontal blue lines.}
\label{fig:stress_point_original_corrected}
\end{figure}

Figure~\ref{fig:stress_point_original_corrected} illustrates the effect of correcting the stress points using INLA-RF2. Specifically, it compares the temporal field estimates at the 100 stress points under the benchmark INLA-based model (black dots and bars denote the posterior means and 95\% credible intervals, respectively) and under the INLA-RF2 hybrid model (red dots and bars). The true simulated values are shown as blue horizontal lines. Overall, applying INLA-RF2 leads to a clear improvement in the temporal field estimates at the selected stress points. The corrected estimates (in red) align more closely with the true values (in blue), whereas the benchmark INLA-based model (in black) shows noticeable bias, especially at time points characterized by rapid level changes. These results highlight the ability of INLA-RF2 to effectively correct the latent field in targeted regions.

Table~\ref{Tab:metrics_alg2} presents the performance measures for both the full dataset and the subset corresponding to the 100 stress points when comparing INLA-RF2 against the benchmark INLA-based model. The INLA-RF2 approach demonstrates clear improvements in accuracy, especially at the stress points, where the RMSE decreases substantially from 0.93 (INLA) to 0.22 (INLA-RF2), and MAE from 0.65 to 0.13. Table~\ref{Tab:metrics_alg2} also highlights a clear difference in uncertainty quantification between the full dataset and the subset of stress points. At the stress points, INLA-RF2 substantially improves coverage, with CP increasing from 0.47 to 0.92, although this gain is accompanied by a wider interval width (AIW increasing from 0.71 to 0.92). Over the full dataset, instead, INLA-RF2 yields narrower intervals than the benchmark INLA-based model (AIW decreasing from 0.69 to 0.52), but with a reduction in coverage (CP decreasing from 0.83 to 0.70). 



\begin{table}[ht]
\centering
\begin{tabular}{lcccccccc}
\toprule
& \multicolumn{2}{c}{RMSE} & \multicolumn{2}{c}{MAE} & \multicolumn{2}{c}{CP} & \multicolumn{2}{c}{AIW} \\
\cmidrule(lr){2-3} \cmidrule(lr){4-5} \cmidrule(lr){6-7} \cmidrule(lr){8-9}
\textbf{Model} & \textbf{Full} & \textbf{Stress} & \textbf{Full} & \textbf{Stress} & \textbf{Full} & \textbf{Stress} & \textbf{Full} & \textbf{Stress} \\
\midrule
INLA     & 0.34 & 0.93 & 0.22 & 0.65 & 0.83 & 0.47 & 0.69 & 0.71 \\
INLA-RF2 & 0.27 & 0.22 & 0.20 & 0.13 & 0.70 & 0.92 & 0.52 & 0.92 \\
\bottomrule
\end{tabular}
\caption{Performance metrics for the INLA-RF2 algorithm and the benchmark INLA-based model, reported for the full dataset and for the subset corresponding to the 100 stress points.}
\label{Tab:metrics_alg2}
\end{table}

\section{Real case study: the Agrimonia dataset}\label{sec:realcasestudy}
The Agrimonia dataset\footnote{\url{https://zenodo.org/records/7956006}} is used for a real-world comparative analysis. This data set comprises daily spatio-temporal observations of multiple air pollutant concentrations, together with covariates concerning weather, emissions, livestock, and land use, for the Lombardy region (Northern Italy) and a 30 km buffer zone around its administrative boundaries. Overall, it includes data for 141 monitoring stations and covers the period from 2016 to 2021 \citep{fasso2023agrimonia}.

\begin{figure}[ht]
    \centering
    \includegraphics[width=\linewidth]{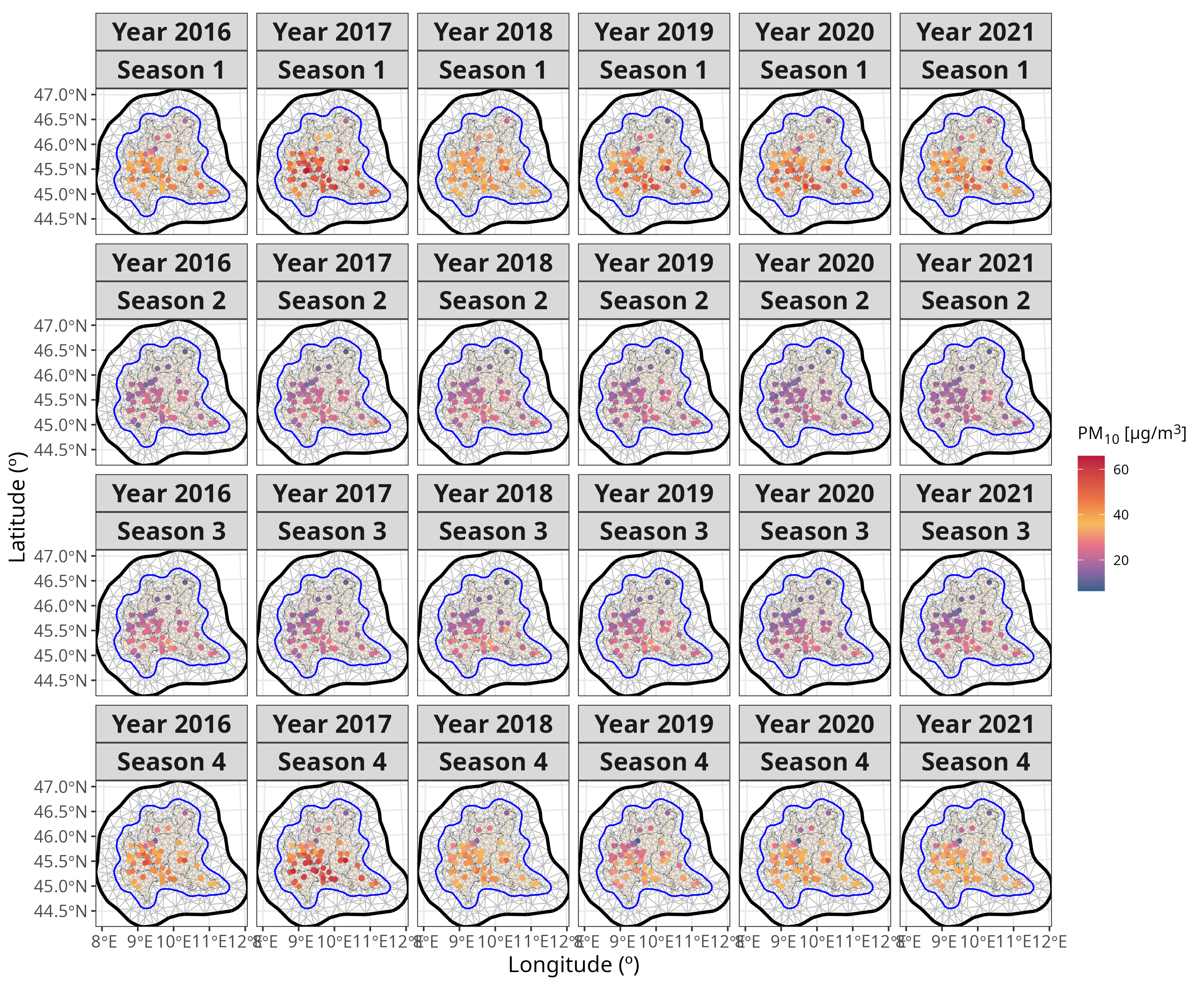}
    \caption{PM$_{10}$ values aggregated at seasonal level for all the available years (2016-2021), and the mesh used in the implementation of the INLA-SPDE approach.}
    \label{fig:Agrimonia_seasons}
\end{figure}

In the present application, PM$_{10}$ concentration was selected as the response variable because of its relevance for air-quality assessment and human health \citep{FIORAVANTI2021}. As explanatory variables, we considered temperature measured at 2 metres above ground level, wind speed measured at 10 metres above ground level, total precipitation, and relative humidity. The analysis was restricted to the subset of 63 monitoring stations located within Lombardy. To simplify the empirical comparison, the daily observations were aggregated at the seasonal level (Jan.-Mar., Apr.-Jun., Jul.-Sep., Oct.-Dec.), see Figure~\ref{fig:Agrimonia_seasons}. 


The benchmark INLA-based spatio-temporal model we fit is defined as follows:
\begin{equation}\label{Eq:agrimonia_1}
\mathbf y \sim \text{MVN}(\boldsymbol\mu, \tau_\ell^{-1}\mathbf I),
\end{equation}
with
\begin{equation}\label{Eq:agrimonia_2}
\boldsymbol\mu = \beta_0 \mathbf 1 + \sum_{j=1}^{4}\beta_j \mathbf z_j + \boldsymbol\xi_{\text{year}} + \boldsymbol\xi_{\text{season}},
\end{equation}
where $\mathbf z_j$ denotes the vector of observations for the $j^{th}$ covariate, $\boldsymbol\xi_{\text{year}}$ and $\boldsymbol\xi_{\text{season}}$ are two spatio-temporal random effects defined at the yearly and seasonal levels, respectively. They are modeled as Gaussian Markov Random Field \citep[GMRF,][]{Rue_2005_GMRF} with precision matrices given by
\begin{equation*}
\mathbf Q_{\text{year}} = \mathbf Q_{\text{AR1,year}} \otimes \mathbf Q_{\text{SPDE}}
\qquad
\mathbf Q_{\text{season}} = \mathbf Q_{\text{AR1,season}} \otimes \mathbf Q_{\text{SPDE}}
\end{equation*}
where $\otimes$ is the Kronecker product, $\mathbf Q_{\text{SPDE}}$ is the SPDE-based spatial precision matrix, while $\mathbf Q_{\text{AR1,year}}$ and $\mathbf Q_{\text{AR1,season}}$ correspond to first-order autoregressive processes over years and seasons (the latter defined as cyclic), respectively \citep{SPDEbook, Virgilio_2020_INLAbook}.

The prior distributions used for the fixed effects $(\beta_0, \beta_1,\beta_2,\beta_3,\beta_4)$ were vague Gaussian priors, $\mathrm{N}(0, \sigma^2 = 1/0.001)$. For the hyperparameters of the spatio-temporal components, we used PC priors. In particular, for the seasonal spatio-temporal component they were specified as $P(\sigma_{\text{season}}>1)=0.5$, $P(\rho_{\text{SPDE,season}}<\rho_0)=0.5$ and $P(\rho_{\text{AR1,season}}>0.5)=0.5$, and for the yearly spatio-temporal effect as $P(\sigma_{\text{year}}>1)=0.5$, $P(\rho_{\text{SPDE,year}}<\rho_0)=0.5$ and $P(\rho_{\text{AR1,year}}>0.5)= 0.5$ (see  Section~\ref{Sec:sim1} for the definition of $\rho_0$).

As in the previous simulation studies, we consider the following models for comparison: the benchmark INLA-based model (see Eq.~\eqref{Eq:agrimonia_1} and \eqref{Eq:agrimonia_2}), the standard RF model, and the two proposed hybrid approaches INLA-RF1 (with and without uncertainty) and INLA-RF2. A primary challenge in applying INLA-RF2 to the real-world case lies in defining a robust procedure for selecting stress nodes. This difficulty stems from the presence of dual spatio-temporal components (seasonal and yearly) and the lack of a general criterion for node selection when dealing with off-node observations and varying data density, including nodes with no directly linked observations. Consequently, in this study the stress nodes for the seasonal component were identified as those associated with observations exhibiting the highest local deviance under the benchmark INLA-based model.
The results obtained following an 80\%-20\% training-test split are reported in Table~\ref{tab:TrainTest_AGRI_Measures_Performance}, while the outcomes regarding the structured spatio-temporal CV are available in \ref{Sec:realcase_details}. 

\begin{table}[ht]
\centering
\begin{tabular}{lcccccccc}
\toprule
& \multicolumn{2}{c}{RMSE} & \multicolumn{2}{c}{MAE} & \multicolumn{2}{c}{CP} & \multicolumn{2}{c}{AIW} \\
\cmidrule(lr){2-3} \cmidrule(lr){4-5} \cmidrule(lr){6-7} \cmidrule(lr){8-9}
\textbf{Model} & \textbf{Train} & \textbf{Test} & \textbf{Train} & \textbf{Test} & \textbf{Train} & \textbf{Test} & \textbf{Train} & \textbf{Test} \\
\midrule
INLA        & $3.04$ & $3.47$ & $2.31$ & $2.63$ & $0.66$ & $0.65$ & $5.24$ & $5.78$ \\
RF          & $3.68$ & $3.68$ & $2.71$ & $2.77$ & $0.94$ & $0.93$ & $14.42$ & $14.42$ \\
INLA-RF1.1  & $2.11$ & $2.54$ & $1.56$ & $1.86$ & $0.72$ & $0.72$ & $4.15$ & $4.70$ \\
INLA-RF1.2  & $2.62$ & $2.86$ & $1.95$ & $2.15$ & $0.94$ & $0.94$ & $10.65$ & $11.00$ \\
INLA-RF2  & $3.01$ & $3.45$ & $2.30$ & $2.62$ & $0.68$ & $0.67$ & $5.39$ & $5.93$ \\
\bottomrule
\end{tabular}
\caption{Performance metrics based on an 80\%-20\% training-test split for the Agrimonia data. The INLA label corresponds to the benchmark INLA-based model with no hybridization, while RF denotes the standard Random Forest. INLA-RF1.1 denotes the INLA-RF1 model without uncertainty propagation, whereas INLA-RF1.2 includes uncertainty propagation.}
\label{tab:TrainTest_AGRI_Measures_Performance}
\end{table}

The results from the training-test split reported in Table~\ref{tab:TrainTest_AGRI_Measures_Performance} show that INLA-RF1.1 and INLA-RF1.2 outperform both the benchmark INLA-based model and standard RF in terms of RMSE and MAE. In particular, INLA-RF1.1 also improves CP relative to the benchmark INLA model, while yielding a lower AIW. INLA-RF1.2 achieves substantially higher coverage than the benchmark INLA model, but at the cost of wider predictive intervals, as reflected by the larger AIW. Although RF provides higher CP than INLA-RF1.1 and nearly identical CP to INLA-RF1.2, this comes with much larger uncertainty, with AIW values that are considerably higher than those of both hybrid approaches. By contrast, INLA-RF2 does not provide a substantial improvement over the benchmark INLA-based model. This is likely because the real data do not exhibit clear localized features requiring node-level correction, and the benchmark model already adapts well to the seasonal and yearly spatio-temporal structure. The slight increase in CP obtained by INLA-RF2 is mainly associated with the larger uncertainty introduced by the extended latent field used for stress-node correction, as reflected in the AIW. These findings are consistent with the results obtained under the structured spatio-temporal CV reported in \ref{Sec:realcase_details}.

\section{Discussion and conclusion}\label{sec:discussion}
In this paper, we proposed two hybrid algorithms, INLA-RF1 and INLA-RF2, for combining a Bayesian spatio-temporal model fitted via the INLA-SPDE approach with Random Forest. The two approaches differ in their objectives and in the way the RF information is incorporated into the model.

INLA-RF1 acts as a post-hoc correction of the mean, by leveraging the RF residual structure and feeding it back into the Bayesian model through an offset. This strategy is simple to implement and proves effective in improving predictive accuracy, especially in the presence of non-linear relationships that are not captured by the large-scale component. The version without uncertainty propagation leads to sharper predictions, while the version with uncertainty propagation provides better calibrated uncertainty at the cost of wider predictive intervals.

INLA-RF2, instead, introduces a localized correction at the level of selected latent field nodes. This approach could be particularly effective in settings characterized by abrupt changes or discontinuities. The results from the simulation study show that INLA-RF2 substantially improves estimation at the selected stress points, although this local gain may come with a trade-off in uncertainty quantification, as reflected by wider credible intervals. However, in the real-world application, neither the seasonal nor the yearly spatio-temporal components exhibited nodes requiring clear correction; consequently, the implementation of INLA-RF2 yielded no significant improvement. This scenario illustrates that identifying specific nodes for local correction can be significantly more complex than refining the linear predictor. In contrast, INLA-RF1, which accounts for non-linear and complex dependencies among explanatory variables via the RF algorithm, offers a more straightforward path to capturing residual structure.


From a practical perspective, the choice between INLA-RF1 and INLA-RF2 depends on the modeling objective. INLA-RF1 is preferable when the goal is to improve overall predictive performance with minimal additional modeling complexity. INLA-RF2 is more suitable when the interest lies in correcting specific regions of the latent field, for example in the presence of structural breaks or localized misspecification. However, its implementation requires careful selection of the nodes to be corrected, which represents an additional modeling step.
Potential strategies for node selection involve first identifying the model component requiring correction and then applying diagnostics based on observational information and model uncertainty. For instance, posterior variance, local RMSE, or local deviance may provide useful quantitative criteria for identifying the most suitable nodes for correction.

Both approaches benefit from the computational efficiency of the INLA-SPDE framework. The use of a Kullback-Leibler divergence-based stopping criterion provides a principled and efficient way to control the iterative procedure, avoiding unnecessary iterations and reducing the risk of overfitting.

The empirical results from the simulation studies and the real-world air quality application indicate that the proposed hybrid approaches can improve predictive accuracy, particularly in challenging spatio-temporal settings. The proposed framework is flexible and can be extended in several directions. For instance, alternative machine learning models could be used in place of RF, and more complex likelihoods could be incorporated within the INLA stage. Future work could also investigate automated strategies for selecting the nodes to be corrected in INLA-RF2, as well as extensions to more complex spatio-temporal settings.

\section*{Author Contributions}
The authors (Mario Figueira = MF, Michela Cameletti = MC, Luca Patelli = LP) contributed to the paper as follows: 
\begin{itemize}
\item Conceptualization: MC, MF, LP
\item Methodology: MC, MF
\item Code: MF, LP
\item Formal data analysis: MF 
\item Writing - original draft: MC, MF, LP
\item Writing - review \& editing: MC, MF, LP 
\item Supervision: MC
\end{itemize}

\section*{Funding}
\begin{sloppypar}
MF acknowledges support from Grant PID2022-136455NB-I00, funded by Ministerio de Ciencia, Innovación y Universidades of Spain (MCIN/AEI/10.13039/501100011033/FEDER, UE) and the European Regional Development Fund.
\end{sloppypar}

\section*{Code and data availability}
The code for generating the data for the two simulation studies, as well as for implementing the proposed algorithms, is publicly available on GitHub at the following repository: \url{https://github.com/LucaPate/INLA-RF/tree/main}.

\section*{Declaration of generative AI and AI-assisted technologies in the writing process}
During the preparation of this work the authors used \cite{openai2025chatgpt} in order to improve the readability and language of the manuscript. After using this tool, the authors reviewed and edited the content as needed and take full responsibility for the content of the published article.

\bibliographystyle{elsarticle-harv}
\bibliography{References}

\appendix
\newpage

\section{Review of INLA-SPDE and RF}\label{sec:reviewmodels}
\subsection{The INLA-SPDE approach}\label{sec:inlaspde}
INLA is a deterministic approximate Bayesian inference approach grounded in the properties of the Gaussian Markov Random Field (GMRF) and its computational efficiency \citep{Rue_2005_GMRF}. This methodology was developed by \cite{Rue_2009_INLA}, implemented in \texttt R through the \texttt {R-INLA} package (\url{https://www.r-inla.org/}), applied in many contexts \citep{Lindgren_2015_INLASpatialReview, Bakka_2018_INLAreview, BlangiardoCameletti2015, SPDEbook, Virgilio_2020_INLAbook} and later extended to more flexible modeling structures \citep{Niekerk_2024_lowrankVB}. This approach allows Bayesian inference to be performed on a wide range of structured additive models, known as Latent Gaussian Models (LGMs). While INLA primarily focuses on computing marginal posterior distributions, it effectively addresses the challenges of determining the joint posterior distributions of interest through a series of nested approximations or a low-rank variational Bayes correction for the latent field \citep{Niekerk_2023_INLAv2, Niekerk_2024_lowrankVB}. This technique enables the computation of marginal likelihoods, standard goodness-of-fit metrics (DIC, WAIC, CPO), cross-validation checks, and posterior predictive distributions.

Since INLA is based on GMRF principles, it offers inherent computational efficiency when modeling data with conditional independence. Generally speaking, given the vector of $n$ observations  $\mathbf{y}=(y_1, \ldots, y_n)$, the likelihood, assuming conditional independence, is described by the following expression:
\begin{equation}\label{eq:likelihood}
\pi(\mathbf{y}|\boldsymbol\eta,\boldsymbol\theta)=\prod_{i=1}^n \pi(y_i \mid \eta_i,\boldsymbol\theta),
\end{equation}
%
where $\eta_i$ is the $i^{th}$ component of the linear predictor vector $\boldsymbol\eta =(\eta_1, \ldots, \eta_n)$, including fixed and random effects, and is defined\footnote{Given that we assume the Gaussian distribution for the response variable it results that $E(y_i)=\eta_i$, i.e., the identity link function is used.} as follows \citep{Lindgren_2015_INLASpatialReview}:
\begin{equation}\label{eq:linearpredictionv2}
\eta_i = \sum_k h_k(z_i^k),
\end{equation}
where $h_k(\cdot)$ is known as \textit{mapping function}. In the case of a fixed effect,  $z_i^k$ is given by the covariate value included in the linear predictor through a linear function, i.e., $h_k(z_i^k)=\beta_k z_i^k$. In the case of a random effect, $z_i^k$ defines the unit index (e.g., the grouping, spatial or temporal index) and $h_k(z_i^k)= u_k(z_i^k)$, where $u_k$ is the value of the latent Gaussian variable assumed for the random effect (with a structured or unstructured prior distribution). The latent field $\mathbf{x}$ encompasses all the latent Gaussian variables which are part of the linear predictor, including the vector of fixed effect coefficients $\boldsymbol{\beta}$ used to include linear effects of the covariates. It is assumed that the latent field $\mathbf{x}$ follows a multivariate normal prior distribution with mean $\boldsymbol\mu$, usually equal to $\mathbf{0}$, and precision matrix $\mathbf{Q}(\boldsymbol\theta_2)$. Given the conditional independence property, it follows that this precision matrix is sparse and this allows for computationally efficiency when performing matrix operations \citep{Rue_2005_GMRF}. 
The vector of hyperparameters is given by $\boldsymbol\theta = (\boldsymbol\theta_1, \boldsymbol\theta_2)$, where $\boldsymbol\theta_1$ refers to the likelihood hyperparameters and $\boldsymbol\theta_2$ to the latent field hyperparameters.

The INLA methodology exploits the assumptions of the model to produce a numerical approximation to the posterior distribution of interest, i.e., the joint posterior distribution of the hyperparameters $ \pi(\boldsymbol\theta \mid \mathbf{y})$ and the latent field marginal distribution $ \pi(x_i \mid \mathbf{y})$.

In Eq.~\eqref{eq:linearpredictionv2} each term $h_k(z_i^k)$ represents the contribution to the linear predictor of a component (also named node) of the latent field $\mathbf{x}$. As explained in \cite{Lindgren_2015_INLASpatialReview}, this formulation only allows each observation to be linked to one element from each $h_k(\cdot)$ effect. This can be a limitation when each observation depends on multiple components of $\mathbf{x}$, as it is the case of spatial random effects. To deal with this, the projector matrix $\mathbf{A}$ is introduced as the design matrix of the model, incorporating covariate values, factor structures and their levels, and weights associated with random effects (e.g., basis weights for spatial effects); it basically defines how each observation is linked to the latent field components. Eq.~\eqref{eq:linearpredictionv2} can thus be rewritten as
\[
\eta_i = \sum_{j=1}^{J}\mathbf{A}_{ij} x_j,
\]
where $\mathbf{A}_{ij}$ is the weight of the $j^{th}$ component of the latent field contributing to the linear predictor $\eta_i$ for the $i^{th}$ observation. Note that the total number of latent components $J$ depends on the model structure. In matrix notation we have that $\boldsymbol\eta = \mathbf{A} \, \mathbf{x}$. In the paper we use the notation LGM($\mathbf{A}, \mathbf{x},\boldsymbol\theta)$ to identify an unique LGM model structure.

In the INLA framework, the term \emph{node} refers to a component of the latent vector $\mathbf x$ (e.g., regression coefficients, temporal effects, or discretized spatial field weights at mesh vertices). For the INLA-RF2 algorithm described in Section~\ref{Sec:INLARF2}, we select a subset of indices $\mathcal K$ (also referred to as stress points or locations) and denote by $\mathbf x_c=\{x_k: k\in\mathcal K\}$ the corresponding components of the latent field. Let $\mathbf S_{\mathcal K}$ be the column-selection matrix that extracts $\mathbf x_c$ from $\mathbf x$ (so that $\mathbf x_c=\mathbf S_{\mathcal K}^\top\mathbf x$). The associated projector matrix used in INLA-RF2 can then be written as $\mathbf A_c=\mathbf A\,\mathbf S_{\mathcal K}$, i.e., $\mathbf A_c$ contains the columns of $\mathbf A$ corresponding to the selected latent components.




The \texttt {R-INLA} package implements also the Stochastic Partial Differential Equation (SPDE) technique, introduced by \cite{Lindgren_2011_SPDE} for modeling point-referenced spatial data in a computationally efficient way (see also \citealt{Lindgren_2015_INLASpatialReview, SPDEbook, Lindgren:spde2022, BlangiardoCameletti2015}).
This method exploits the fact that the solution to a specific SPDE is a Gaussian Field (GF) with Matérn covariance function (see Eq.~\eqref{eq:matern} for its definition). The novelty of the INLA-SPDE approach is to represent the Matérn GF as a GMRF by using the finite element method, i.e. by defining the Matérn field as a linear combination of basis functions defined on a triangulation (i.e., mesh) of the considered spatial domain (see for example Figure \ref{fig:simulateddata}). This eliminates the need to work with a dense covariance matrix and instead allows the use of a sparse precision matrix, avoiding computationally expensive inversion processes. Note that when applying the SPDE approach in a two-dimensional spatial domain, the Matérn smoothness parameter $\nu$ is set equal to one and the spatial range is given by $\rho=\sqrt{8\nu}/\kappa$ and interpreted as the distance at which the correlation drops to approximately $0.13$ \citep[see][]{Lindgren_2011_SPDE}.

The linear combination of basis functions used to approximate the Matérn GF $\xi(\mathbf s)$ is given by
\begin{equation}\label{eq:GFapproximation}
\xi(\mathbf s)=\sum_{g=1}^G \psi_g(\mathbf s)\tilde\xi_g,
\end{equation}
where $G$ is the total number of mesh vertices, $\psi_g(\cdot)$ are piecewise linear basis functions, and $\tilde{\bm\xi}=(\tilde \xi_1, \ldots, \tilde\xi_G)$ are latent weights modeled as a multivariate zero-mean Gaussian random variable with a sparse precision matrix derived from the SPDE formulation (that's why $\tilde{\mathbf \xi}$ is a GMRF). When implementing the SPDE approach with INLA, the summation in Eq.~\eqref{eq:GFapproximation} is used as one of the components of the linear predictor defined in Eq.~\eqref{eq:linearpredictionv2}, as an indexed random effect. Accordingly, the latent Gaussian vector $\tilde{\mathbf \xi}$ will be included in the INLA latent field $\mathbf{x}$  \citep{Lindgren_2015_INLASpatialReview}. In this case the projector matrix $\mathbf{A}$ (with dimension $n\times G)$ maps the GMRF $\tilde{\mathbf \xi}$ from the $G$ mesh vertices to the $n$ observation locations such that $\mathbf A_{ig}=\psi_g(\mathbf s_i)$. In particular, when a spatial location $\mathbf{s}_i$ lies within a triangle defined by three mesh vertices, the corresponding row of $\mathbf{A}$ contains three nonzero elements, which sum to one \citep{SPDEbook}.

\subsection{Random Forest}\label{sec:RF}
Random Forest (RF, \citealp{breiman2001random}) is a widely adopted non-parametric ML model which has gained notoriety given its simple implementation and the satisfactory predictive results. 
The RF algorithm is based on two main elements: bootstrap \citep{efron1979boostrap} and decision trees \citep{breiman1984classification}. The two methods interact in the following way: a large number of bootstrapped samples are generated from the available data, and a decision tree is fitted on each of them.
Decision tree growth is an iterative process in which the predictor space is partitioned at each step until a stopping rule is satisfied. The splitting optimization algorithm determines the best predictor and corresponding cut off value to minimize local residual errors. In the specific case of RF, the optimal predictor is selected from a randomly chosen subset of the original variables (see, e.g., \citealp{james2021introduction}). Being based on decision trees, RF is characterized by high flexibility in modeling the relationship between predictors and the response variable, being able to capture non-linear relationships and interactions between and within predictors.

A limitation of standard RF in applications involving correlated data is that it does not explicitly account for spatial and/or temporal dependence between observations unless this information is incorporated through dedicated modifications or additional predictors \citep{Patelli2024path, saha2023random}. This means that RF can be effectively used only to estimate the LS component of Eq.~\eqref{Eq:generalmodel}. This limitation arises from the fact that the RF bootstrap does not consider, when creating samples, information coming from the spatial location or temporal indexing of the observations. Additionally, in the absence of spatial and/or temporal information introduced through predictors, the splitting mechanism cannot leverage them. 
In \cite{Patelli2024path}, a comprehensive literature review was conducted to identify the adopted strategies to make RF spatially aware, and these strategies were then systematically organized following a taxonomy. A synthesis of these strategies reveals three main categories, distinguished by when spatial information is incorporated into the RF predictive process (at the covariates level, modifying the internal algorithm or as a post-processing phase).
Among the identified contributions, RF-GLS \citep{saha2023random} appears to be the most promising because it incorporates spatial correlation during the tree growth. However, it appears to be computationally intensive \citep{heaton_millane_rhodes_2025} and constrained to the case of purely spatial or temporal dependency structure. Consequently, it is not suitable for spatio-temporal applications.
Another limitation of RF is that, being a non-parametric method, it lacks direct interpretability and uncertainty evaluation of the predictions. 

\section{Performance metrics and stopping criteria}\label{AppendixB:metrics}

\subsection{Predictive performance metrics}
We define here the predictive performance metrics used to compare models. We denote by $y_i$ the observed value and by $\hat y_i$ the corresponding prediction.

\begin{enumerate}
    \item \textit{Root mean square error} ($RMSE$) - a measure of the average magnitude of prediction error, giving higher weight to larger errors:
    $$
    RMSE = \sqrt{\frac{1}{n}\sum_{i=1}^n (y_i - \hat{y}_i)^2}.
    $$

    \item \textit{Mean absolute error} ($MAE$) - the average of the absolute differences between predicted and observed values:
    $$
    MAE = \frac{1}{n}\sum_{i=1}^n |y_i - \hat{y}_i|.
    $$

    \item \textit{Coverage probability} ($CP$) - the proportion of observed values that fall within the predictive interval defined by the $0.025$ and $0.975$ quantiles (i.e., the $95\%$ credible interval). It assesses the calibration of uncertainty estimates and is defined as
    $$
    CP = \frac{1}{n}\sum_{i=1}^n \mathbb{I}\left(F^{-1}_{\hat{y}_i}(p_l) \leq y_i \leq F^{-1}_{\hat{y}_i}(p_u)\right),
    $$
    where $\mathbb{I}(\cdot)$ is the indicator function, which is $1$ if the condition is true and $0$ otherwise, $F^{-1}_{\hat{y}_i}(\cdot)$ is the quantile function of the posterior predictive distribution of $\hat{y}_i$, and $p_l$ and $p_u$ are the lower and upper quantile probabilities ($0.025$ and $0.975$, respectively).

    \item \textit{Average interval width} ($AIW$) - the average width of the the $95\%$ credible interval. It reflects the overall uncertainty in the predictions, with narrower intervals indicating higher precision. It is given by
    $$
    AIW = \frac{1}{n}\sum_{i=1}^n \left(F^{-1}_{\hat{y}_i}(p_u) - F^{-1}_{\hat{y}_i}(p_l)\right).
    $$
\end{enumerate}

\subsection{Stopping criterion}\label{sec:KLD}
The value of the Kullback-Leibler divergence $D_{KL}$ can be computed using two different approaches: (1) using the KLD given by the conditional posterior distribution of the latent field in the modal configuration of the hyperparameters, or (2) using the KLD given by the marginal posterior distribution of the latent field. In the following, the notation $KLD(\cdot || \cdot)$ is used to denote the KLD between two distributions.

The first approach accounts for the ``average'' Kullback-Leibler divergence between two multivariate distributions, i.e., the KLD accounts for the dimension of the latent field. In the INLA methodology the joint conditional posterior distribution $\pi(\mathbf{x} \mid \mathbf{y}, \boldsymbol\theta)$ is computed as a Gausssian approximation, which allows us to compute the $D_{KL}$ between two consecutive iteration of the algorithm ($i-1$ and $i$) at a low computational cost as follows:
\begin{equation}
    \begin{split}
            D_{KL} & =  \displaystyle{KLD}\left(\pi(\mathbf{x} \mid \mathbf{y}^{(i-1)}, \boldsymbol\theta^{(i-1)}) \mid \mid \pi(\mathbf{x} \mid \mathbf{y}^{(i)}, \boldsymbol\theta^{(i)})\right) \\[1.5mm]
     & =  \displaystyle \frac{1}{2} \left( (\boldsymbol\mu^{(i)} - \boldsymbol\mu^{(i-1)})^{\top} \mathbf{Q}^{(i)} (\boldsymbol\mu^{(i)} - \boldsymbol\mu^{(i-1)}) + \text{tr}(\mathbf{Q}^{(i)}{\mathbf{Q}^{(i-1)}}^{-1}) - \log\frac{|\mathbf{Q}^{(i)}|}{|\mathbf{Q}^{(i-1)}|} \right) \; ,
     \label{eq:averageKLD}
    \end{split}
\end{equation}
where $\top$ is the transpose operation, $\text{tr}$ is the matrix trace and $|\cdot|$ the determinant. Moreover, $\pi(\mathbf{x} \mid \mathbf{y}^{(i)}, \boldsymbol\theta^{(i)})$ is the conditional posterior distribution at the $i^{th}$ iteration. Therefore, $\pi(\mathbf{x} \mid \mathbf{y}^{(i)}, \boldsymbol\theta^{(i)})$ follows a multivariate Gaussian distribution denoted by $\text{MVN}(\boldsymbol\mu^{(i)}, \mathbf{Q}^{(i)})$, with mean vector $\boldsymbol\mu^{(i)}$ and precision matrix $\mathbf{Q}^{(i)}$.

The second approach is computed as average difference in the marginal posterior distributions of all the latent field components (or nodes) represented by $\pi(x_j \mid \mathbf{y}^{(i-1)})$: 
\begin{equation}
D_{KL} = \frac{1}{J} \sum_{j=1}^{J} \displaystyle{KLD}\left(\pi(x_j \mid \mathbf{y}^{(i-1)}) \mid \mid \pi(x_j \mid \mathbf{y}^{(i)})\right),
\label{eq:averageKLD2}
\end{equation}
with $J$ being the total number of latent field components. 
 
Alternatively, instead of using an average measure we can consider the maximum KLD computed over the set of latent field nodes denoted by $\mathcal{J}$. This measure is defined as follows: 
\begin{equation}
D_{KL} = \max_{j \in \mathcal{J}}\left\{\displaystyle{KLD}\left(\pi(x_j \mid \mathbf{y}^{(i-1)}) \mid \mid \pi(x_j \mid \mathbf{y}^{(i)})\right)\right\}.
\end{equation}
This has the advantage of controlling whether every node update remains below the threshold $\delta$. Note that it can also be computed considering a subset of nodes ($\mathcal{J}' : \mathcal{J}'\subset \mathcal{J}$) if we are interested in evaluating the changes in the marginal distributions only for the nodes in the subset $\mathcal{J}'$.

\section{Details about the simulation studies}\label{Sec:simulations_details}
\subsection{Spatio-temporal simulation study for INLA-RF1}\label{Appendix:sim1data}
\subsubsection{Data simulation}
In the first simulation study, we simulate a dataset with $T=8$ time points and $n=150$ spatial locations per each time location, assuming that the spatial locations are changing in time in the given spatial domain (see Figure~\ref{fig:simulateddata}). The study region and the spatial locations refer to the Paran\'a state and are available in the \texttt {R INLA} package \citep{SPDEbook}. We simulate the data using the following spatio-temporal model which is widely adopted in the environmental modeling literature \citep[see e.g.,][]{Cameletti2013, Moraga2019Geospatial, STwithR2019, FIORAVANTI2021, FIORAVANTI2022, Fioravanti2023, Otto2024PM25Lombardy}. We consider two quantitative predictors ($z_1$ and $z_2$) and one categorical regressor with three classes included using dummy variables ($D_1$, $D_2$ and $D_3$ with coefficients $\gamma_1$, $\gamma_2$ and $\gamma_3$, respectively) which enters in the model through two non-linear functions $f_1(\cdot)$ and $f_2(\cdot)$:

\begin{equation}
\begin{split}
y(\mathbf s_i,t)  &=   f_1\left( z_1(\mathbf s_i,t)\right) +f_2\left( z_2(\mathbf s_i,t)\right)+ \sum_{j=1}^{3} \gamma_j D_j(\mathbf s_i,t) + \omega(\mathbf s_i,t) + \epsilon(\mathbf s_i,t) \\
\omega(\mathbf s_i,t) &= a\omega(\mathbf s_i,t-1)+\xi(\mathbf s_i,t).\label{Eq:simulationmodel}
\end{split}
\end{equation}

The measurement error is assumed to be normally distributed, i.e., $\epsilon(\mathbf s_i,t)\sim \text N(0,\sigma^2_\epsilon)$, while the latent process $\omega(\mathbf s_i,t)$ is characterized by a first order autoregressive temporal dynamics (AR(1)) with coefficient $a$ and by spatially correlated innovations. In particular, $\xi(\mathbf s_i,t)$ is assumed to be a zero-mean Gaussian (isotropic and second order stationary) process with spatio-temporal covariance function given by
\[
Cov\left(\xi(\mathbf s_i, t_k), \xi (\mathbf s_j, t_l)\right) = \begin{cases} 
0, & \text{for } t_k \neq t_l \\ 
\mathcal{C}(h), & \text{for } t_k = t_l
\end{cases},
\]
where $h=||\mathbf s_i - \mathbf s_j||\in \mathbb R$ is the Euclidean distance and $\mathcal C(h)$ is the Mat\`ern spatial covariance function defined as follows \citep{STwithR2019}: 
\begin{equation}\label{eq:matern}
    \mathcal{C}(h)=\sigma^2\frac{2^{1-\nu}}{\Gamma(\nu)}\left(\kappa h \right)^\nu \mathcal{K}_\nu\left(\kappa h\right).
\end{equation}
The term $\sigma^2$ is the spatial variance and $\mathcal{K}_\nu$ is the modified Bessel function of the second kind of order $\nu>0$, the latter being the parameter that controls the smoothness of the spatial process. The term $\kappa>0$ is a scaling parameter related to the spatial range $\rho$, represented by the distance at which the spatial correlation is negligible. 

For the simulation we set $\gamma_1=0.727$, $\gamma_2=-1.027$ and $\gamma_3 = 0.3$. The values of $z_1$ are drawn from a standard Normal distribution with $f_1(\cdot)$ being a non-linear function given by $f_1(z_1)=2z_1\sin(2z_1)$. Furthermore, $z_2$ is simulated from a Uniform distribution defined between 0 and 1 with $f_2(\cdot)$ being another non-linear function defined as $f(z_2)=\sin(z_2^4) + \cos(2.5\pi z_2)$. 

For the spatial covariance function we assume $\sigma^2=1$ and $\rho=3.627$ (given by half the maximum distance for the considered domain). The measurement error $\epsilon$ is assumed to be a mean-zero Gaussian process with variance set equal to $\sigma^2_\epsilon=1/50=0.02$. Finally, for the AR(1) part we assume $a=0.7$.

\subsubsection{Cross-Validation results}\label{AppendixA:CV}
In this section we propose a Cross-Validation (CV) analysis for the spatio-temporal simulation study presented in Section~\ref{Sec:sim1} and \ref{Appendix:sim1data}. The aim is to evaluate the model's performance under a structured spatio-temporal $k$-fold partitioning in order to take into account the spatio-temporal structure of the data when splitting into training and test. The procedure used to construct the spatio-temporal blocks (or clusters) can be summarized as follows: first, the temporal nodes to be aggregated are selected. For each temporal aggregation, a fixed spatial partitioning of the data using the $k$-means algorithm is applied using the spatial coordinates of the observations. This results in a spatio-temporal partition where the overall dimensionality is determined by the number of aggregated temporal nodes and the number of centroids specified in the $k$-means algorithm. For the CV analysis, one of the partitions is selected as the test set, and the remaining partitions are used as the training set. For example, if the data is divided into six spatio-temporal partitions, one partition is used for testing purposes and the remaining partitions serve as the training set. This process is then repeated until all the partitions have been used as the test set. As a result, we obtain performance metrics for each partition, which can then be averaged to compute an overall performance estimate. 

\begin{figure}[ht]
    \centering
    \includegraphics[width=\linewidth]{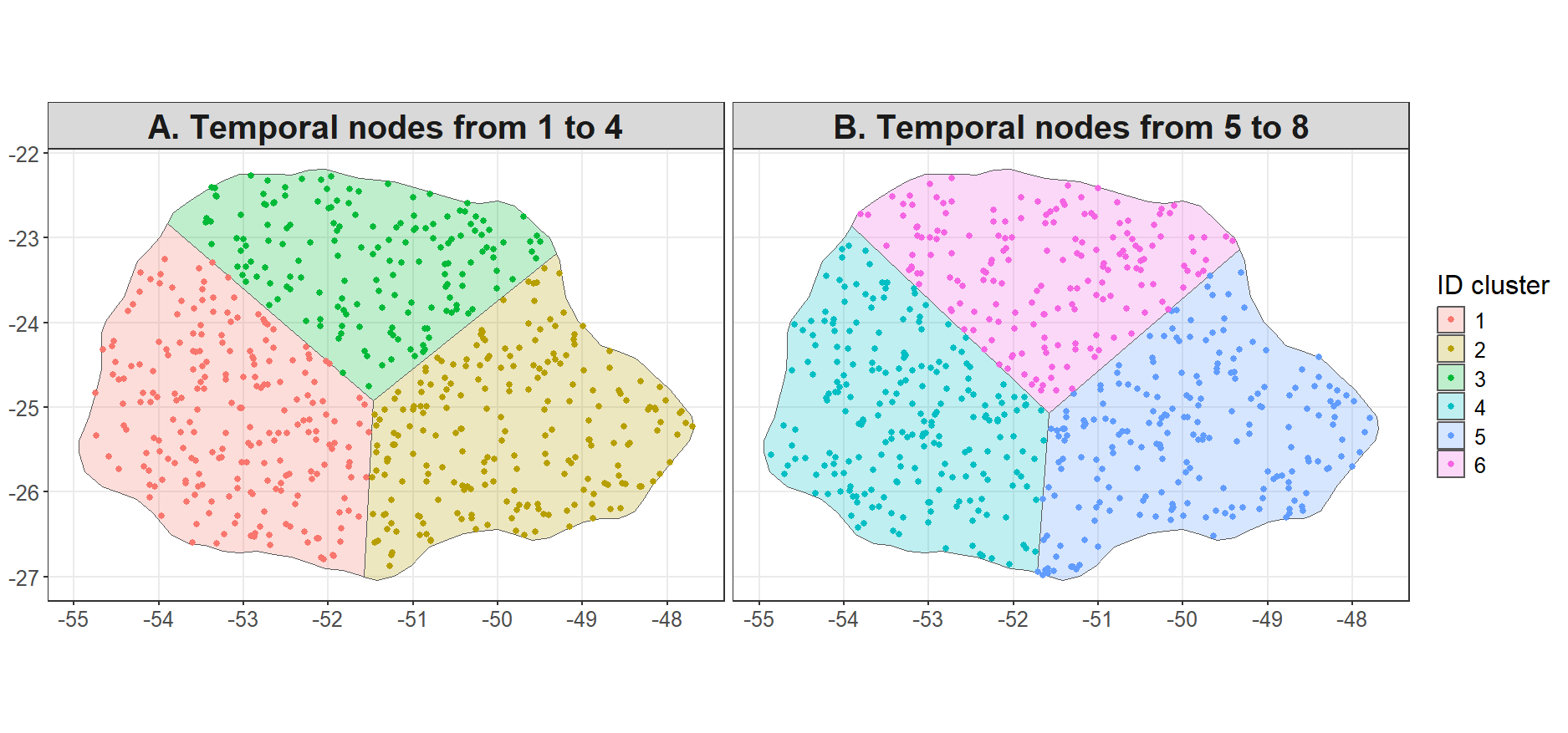}
    \caption{Spatio-temporal clustering with 6 spatio-temporal blocks, showing the data related to each partition.}
    \label{fig:spatio_temporal_cluster_6}
\end{figure}

Figure~\ref{fig:spatio_temporal_cluster_6} shows the six spatio-temporal clusters along with the data assigned to each of them. 
Table~\ref{tab:performance_metrics_CV} shows that the average performance is consistent with the results obtained when using a standard ($80\%-20\%$) training-test split. Specifically, applying INLA-RF1 improves the identification of true values. The most precise predictions are obtained when uncertainty from the RF component is not propagated to the INLA-based model, as expected. When uncertainty is propagated, the predictive intervals become wider than in the benchmark approach. Nevertheless, both versions of INLA-RF1 - whether uncertainty is propagated or not - lead to lower estimation error and better coverage of the true values, although the version with uncertainty propagation results in wider intervals compared to the benchmark method.

\begin{table}[ht]
\centering
\begin{tabular}{lcccccccc}
\toprule
& \multicolumn{2}{c}{RMSE} & \multicolumn{2}{c}{MAE} & \multicolumn{2}{c}{CP} & \multicolumn{2}{c}{AIW} \\
\cmidrule(lr){2-3} \cmidrule(lr){4-5} \cmidrule(lr){6-7} \cmidrule(lr){8-9}
\textbf{Model} & \textbf{Train} & \textbf{Test} & \textbf{Train} & \textbf{Test} & \textbf{Train} & \textbf{Test} & \textbf{Train} & \textbf{Test} \\
\midrule
INLA        & 1.37 & 1.74 & 1.02 & 1.32 & 0.63 & 0.71 & 2.11 & 3.27 \\
RF          & 1.06 & 1.17 & 0.82 & 0.92 & 0.95 & 0.94 & 4.27 & 4.22 \\
INLA-RF1.1  & 0.62 & 1.17 & 0.47 & 0.91 & 0.77 & 0.77 & 1.35 & 2.72 \\
INLA-RF1.2  & 0.62 & 1.17 & 0.47 & 0.90 & 0.97 & 0.90 & 2.80 & 3.68 \\
\bottomrule
\end{tabular}
\caption{Performance metrics using the spatial CV training-test scheme. The INLA label corresponds to the benchmark INLA-based model with no hybridization, while RF denotes the standard Random Forest. INLA-RF1.1 denotes the INLA-RF1 model without uncertainty propagation, whereas INLA-RF1.2 includes uncertainty propagation.}
\label{tab:performance_metrics_CV}
\end{table}


\subsection{Purely temporal simulation study for INLA-RF2}\label{Appendix:sim2data}
\subsubsection{Data simulation}
In this section we provide an example involving a purely temporal application of INLA-RF2 based on simulated data.
The case study is constructed by simulating a time series $\mathbf{y}$ with $n=2000$ time points including $k=10$ jumps introduced at equally spaced time locations. The simulation model is defined as follows using matrix notation:
\begin{equation}
\begin{array}{rcl}
\mathbf{y} \sim \text{MVN}(\boldsymbol\mu, \tau_{\ell}^{-1}\mathbf{I}) \\ 
\boldsymbol\mu =  \beta_0 \mathbf 1  + \mathbf{u} \;,  
\end{array}
\label{eq:simulation_model_structure}
\end{equation}

where $\beta_0$ is the intercept, $\mathbf 1$ is the $n$-dimensional vector of ones, and $\mathbf u$ is a temporal latent effect combining a smooth random-walk component with abrupt jumps. Specifically,
\[
\mathbf{u} = \mathbf{D}\mathbf{w} + \mathbf{u}_r,
\]
where $\mathbf{u}_r$ is a first-order random walk (RW1), meaning that its vector of first-order increments, denoted by $\Delta \mathbf{u}_r$, satisfies
\[
\Delta \mathbf{u}_r \sim \mathrm{MVN}\!\left(\mathbf{0}, \tau_{u_r}^{-1}\mathbf{I}\right).
\]

The matrix $\mathbf D$ is an $n\times k$ cumulative jump-incidence matrix. In particular, let $t_1,\dots,t_k$ denote the jump locations. Then the $(i,j)$ entry of $\mathbf D$ is defined as
\[
D_{ij} = \mathbbm{1}(i \ge t_j),
\]
where $\mathbbm{1}(i \ge t_j)$ equals 1 if $i \ge t_j$ and 0 otherwise. This means that the $j^{th}$ column is equal to 0 before time $t_j$ and 1 from time $t_j$ onward. As a consequence, the product $\mathbf D\mathbf w$ yields a piecewise-constant cumulative jump component, where each element $w_j$ represents the magnitude of the jump occurring at time $t_j$. The jump magnitudes are generated as
\[
\mathbf w = \mathrm{sign}(\mathbf w_s - 0.5)\odot \mathbf w_j,
\]
where \(\odot\) denotes element-wise multiplication, $\mathbf w_s \sim \mathrm{Binom}(1,\boldsymbol\pi_s)$ determines the jump signs and $\mathbf w_j \sim \mathrm{MVN}(\boldsymbol\mu_{w_j},\tau_{w_j}^{-1}\mathbf I)$ determines their absolute sizes. 

In the simulation, the success probabilities in the vector $\boldsymbol{\pi}_s$ are fixed at $0.5$ for all jump locations, the mean jump magnitudes in $\boldsymbol{\mu}_{w_j}$ are set to $5$ for all jumps, and $\tau_{w_j}=100$. The jump locations are equally spaced, with spacing $\lfloor n/(k+1)\rfloor = 181$, so that the jumps occur at times $181, 362, \dots, 1810$. Moreover, we set $\beta_0=2$ and $\tau_\ell=\tau_{u_r}=20$. The simulated data $\mathbf{y}$  are shown in Figure~\ref{fig:simulated_data}, where both the temporal trend and the jumps in the time series are visible.

\section{Details about the real case comparison}\label{Sec:realcase_details}


\subsection{Spatio-temporal cross validation results}\label{Appendix:CV_real_data}
\begin{figure}[ht]
    \centering
    \includegraphics[width=\linewidth]{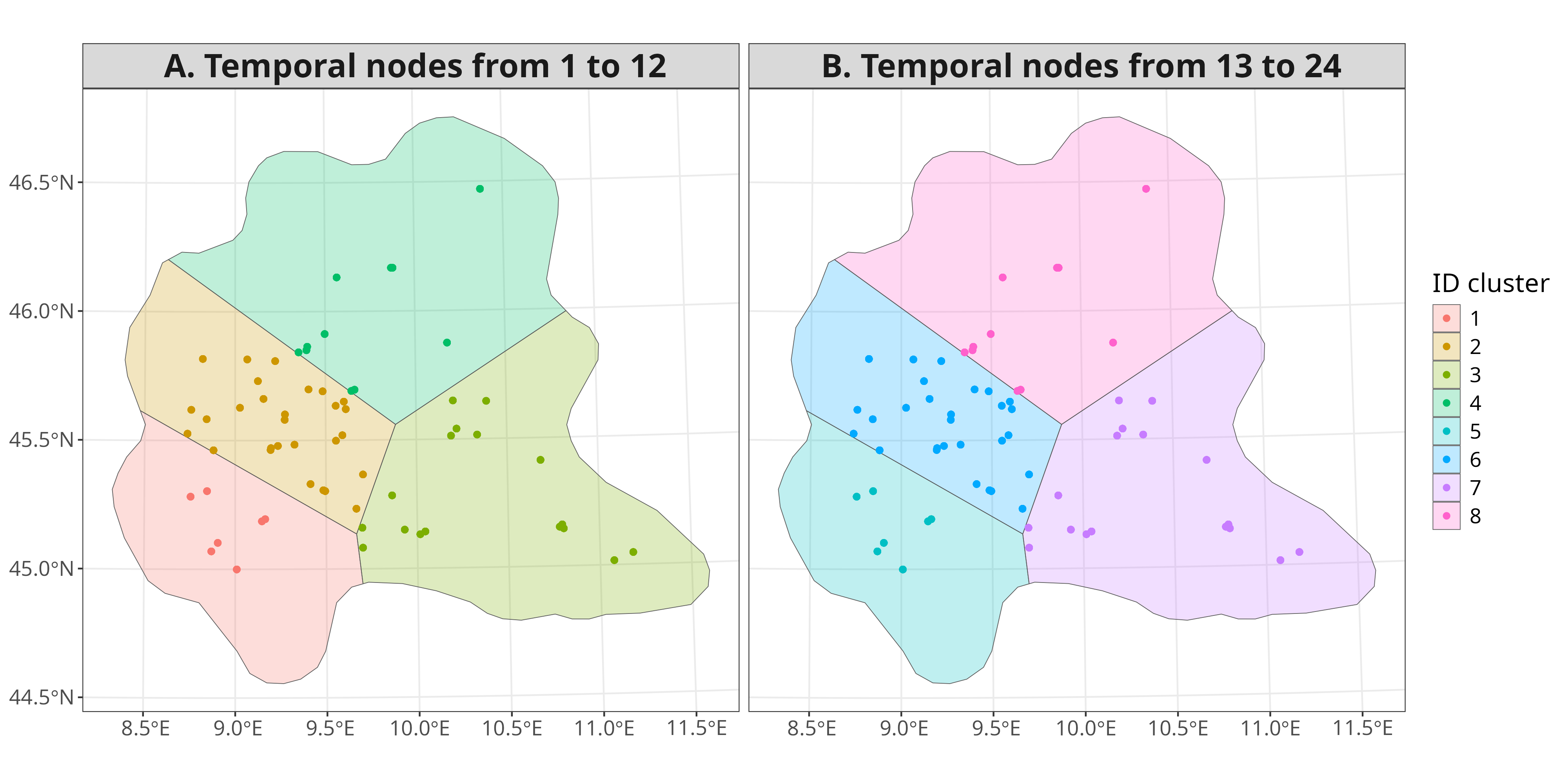}
    \caption{Agrimonia spatio-temporal clustering with 8 spatio-temporal blocks, showing the data related to each partition.}
    \label{fig:Agrimonia_CV}
\end{figure}

Following the same procedure described in \ref{AppendixA:CV}, we carried out a structured spatio-temporal CV analysis for the Agrimonia case study presented in Section~\ref{sec:realcasestudy}. This analysis complements the 80\%-20\% training-test split and aims at evaluating the robustness of the model comparison under partitions that explicitly respect the spatio-temporal structure of the data. The observations were divided into 8 spatio-temporal blocks, illustrated in Figure~\ref{fig:Agrimonia_CV}, and each block was used in turn as test set. Table~\ref{tab:performance_metrics_CV_Agrimonia} reports the average performance metrics across the resulting 8 cross-validation partitions.

\begin{table}[ht]
\centering
\begin{tabular}{lcccccccc}
\toprule
& \multicolumn{2}{c}{RMSE} & \multicolumn{2}{c}{MAE} & \multicolumn{2}{c}{CP} & \multicolumn{2}{c}{AIW} \\
\cmidrule(lr){2-3} \cmidrule(lr){4-5} \cmidrule(lr){6-7} \cmidrule(lr){8-9}
\textbf{Model} & \textbf{Train} & \textbf{Test} & \textbf{Train} & \textbf{Test} & \textbf{Train} & \textbf{Test} & \textbf{Train} & \textbf{Test} \\
\midrule
INLA        & 3.00 & 4.79 & 2.28 & 3.61 & 0.64 & 0.66 & 4.89 & 8.26 \\
RF          & 3.58 & 4.39 & 2.64 & 3.33 & 0.94 & 0.89 & 14.02 & 14.02 \\
INLA-RF1.1  & 2.08 & 3.90 & 1.55 & 2.98 & 0.73 & 0.78 & 4.04 & 8.96 \\
INLA-RF1.2  & 2.26 & 4.07 & 1.68 & 3.13 & 0.95 & 0.89 & 9.58 & 12.60 \\
INLA-RF2  & 2.91 & 4.77 & 2.27 & 3.58 & 0.67 & 0.68 & 5.01 & 8.38 \\
\bottomrule
\end{tabular}
\caption{Mean of the performance metrics along the 8 partitions for the spatio-temporal CV training-test scheme. The INLA label corresponds to the benchmark INLA-based model with no hybridization, while RF denotes the standard Random Forest. INLA-RF1.1 denotes the INLA-RF1 model without uncertainty propagation, whereas INLA-RF1.2 includes uncertainty propagation. Finally, INLA-RF2 refers to the second INLA-RF algorithm, intended for node correction.}
\label{tab:performance_metrics_CV_Agrimonia}
\end{table}

\end{document}